%% file: pereira-nsdi24.tex
\documentclass[letterpaper,twocolumn,10pt]{article}
\usepackage{usenix-2020-09}

\input{preamble}

\hypersetup{
pdflang={en-US},
pdftitle={Automatic Parallelization of Software Network Functions},
pdfsubject={Networking},
pdfkeywords={Maestro, Parallelization, Network Functions},
pdfauthor={%
    Francisco Pereira;
    Fernando M. V. Ramos;
    Luis Pedrosa;
},
}

\begin{document}

\title{\Large \bf Automatic Parallelization of Software Network Functions}

\newcommand*{\ist}{\includegraphics[scale=0.038]{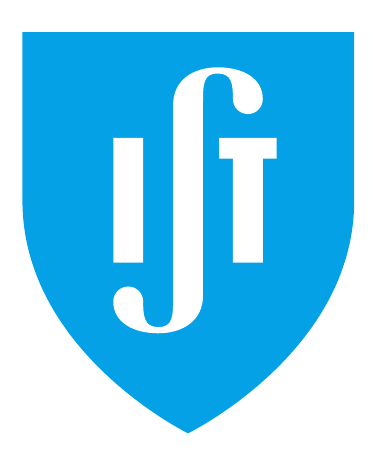}}

\author{
    {\rm Francisco Pereira \ist \quad}
    {\rm Fernando M. V. Ramos\,\ist \quad}
    {\rm Luis Pedrosa\,\ist} \\
    {\small \ist\,INESC-ID, Instituto Superior Técnico, University of Lisbon}
}

\maketitle

\input{sections/0-abstract}
\input{sections/1-introduction}
\input{sections/2-motivation}
\input{sections/3-architecture}
\input{sections/4-implementation}
\input{sections/5-limitations}
\input{sections/6-evaluation}
\input{sections/7-related-work}
\input{sections/8-conclusions}
\input{sections/9-acknowledgments}

\clearpage

\bibliographystyle{plain}
\interlinepenalty=100
\bibliography{pereira-nsdi24}
\interlinepenalty=0


\input{sections/10-appendix}

\end{document}

%% file: preamble.tex
\usepackage{titlesec}
\usepackage{cuted}
\usepackage{mathtools}
\usepackage[ruled,vlined,norelsize,\languagename]{algorithm2e}
\usepackage{soul}
\usepackage[inline]{enumitem}
\usepackage{subfig}

\input{fpmacros}

\newcommand{\dsepA}{\mathbin{\scalebox{1.5}{.}}}

\let\oldnl\nl
\newcommand\nonl{%
  \renewcommand{\nl}{\let\nl\oldnl}}

\SetKw{Continue}{continue}
\SetNlSty{}{}{}

\SetKw{Not}{not}
\SetNlSty{}{}{}

\newcommand{\maestro}{Maestro\xspace}
\newcommand{\librs}{RS3\xspace}

\crefname{lstlisting}{Listing}{Listings}
\Crefname{lstlisting}{Listing}{Listings}

%% file: fpmacros.tex
\usepackage{xspace}
\usepackage{tikz}
\usepackage{url}
\usepackage{hyperref}
\usepackage{amsmath}
\usepackage[noabbrev,capitalise]{cleveref}
\usepackage{listings}
\usepackage{soul}
\usepackage{ifthen}
\usepackage{inconsolata}

\definecolor{fp_color}{RGB}{0, 110, 184} 
\definecolor{changes_color}{RGB}{0, 162, 227} 

\crefformat{section}{\S#2#1#3} 

\input{compress-whitespace}

\definecolor{fp_green}{rgb}{0,0.6,0}
\definecolor{fp_gray}{rgb}{0.5,0.5,0.5}
\definecolor{fp_mauve}{rgb}{0.58,0,0.82}

\lstdefinestyle{C}{ 
  backgroundcolor=\color{white},
  basicstyle=\ttfamily\footnotesize,
  breakatwhitespace=false,
  breaklines=true,
  captionpos=b,
  commentstyle=\color{fp_green},
  deletekeywords={},
  escapeinside={\%*}{*)},
  extendedchars=true,
  frame=single,
  keepspaces=true,
  language=C,
  keywordstyle  = \color{blue},
  keywordstyle  = [2]{\color{teal}},
  keywordstyle  = [3]{\color{magenta}},
  otherkeywords = {pkt_t,port_t,flow_t,bool,uint8_t},
  morekeywords  = [2]{allocate_port,map_get,map_put,forward,drop,rss_configure,map_init,rte_lcore_id},
  morekeywords  = [3]{LAN,WAN,PUBLIC_IP},
  numbers=left,
  numbersep=5pt,
  numberstyle=\tiny\color{fp_gray},
  rulecolor=\color{black},
  showspaces=false,
  showstringspaces=false,
  showtabs=false,
  stringstyle=\color{fp_mauve},
  tabsize=2,
}

\newcommand{\notes}[1]{}
\renewcommand{\notes}[1]{#1}

\newcommand{\etal}[0]{\textit{et al.}\xspace}
\newcommand{\ie}[0]{\textit{i.e.}\xspace}
\newcommand{\eg}[0]{\textit{e.g.}\xspace}
\newcommand{\etc}[0]{\textit{etc.}\xspace}

\DeclareRobustCommand\circled[1]{\tikz[baseline=(char.base)]{
            \node[shape=circle,fill,inner sep=1.5pt] (char)
            {\textcolor{white}{#1}};}}



%% file: compress-whitespace.tex
\usepackage[font=small,skip=0pt]{caption}

\renewcommand{\subsubsection}[1]{\noindent\textbf{#1.}\hspace{1ex}}
\setlist[itemize]{noitemsep,topsep=0pt,after=\vspace{.5\baselineskip}}
\setlist[enumerate]{noitemsep,topsep=0pt,after=\vspace{.5\baselineskip}}

\titlespacing*{\section}{0pt}{2mm}{1mm}  
\titlespacing*{\section}{0pt}{2mm}{1mm}  
\titlespacing*{\subsection}{0pt}{2mm}{1mm}  

\makeatletter
\renewcommand{\@maketitle}{%
     \newpage
     \vspace{-9.75em} 
     \vbox to 2.5in{
        \vspace*{\fill}
        \vskip 2em
        \begin{center}%
            {\Large\bf \@title \par}%
            \vskip 0.375in minus 0.300in
            {\large\it
            \lineskip .5em
            \vspace{-0em} 
            \begin{tabular}[t]{c}\@author
            \end{tabular}\par}%
        \end{center}%
        \par
        \vspace*{\fill}
        \vspace{-6em} 
    }
}
\makeatother

%% file: sections/0-abstract.tex
\begin{abstract}

Software network functions (NFs) trade-off flexibility and ease of deployment for an increased challenge of performance. The traditional way to increase NF performance is by distributing traffic to multiple CPU cores, but this poses a significant challenge: \emph{how to parallelize an NF without breaking its semantics?} We propose \maestro, a tool that analyzes a sequential implementation of an NF and automatically generates an enhanced parallel version that carefully configures the NIC's Receive Side Scaling mechanism to distribute traffic across cores, while preserving semantics. When possible, \maestro orchestrates a shared-nothing architecture, with each core operating independently without shared memory coordination, maximizing performance. Otherwise, \maestro choreographs a fine-grained read-write locking mechanism that optimizes operation for typical Internet traffic. We parallelized 8 software NFs and show that they generally scale-up linearly until bottlenecked by PCIe when using small packets or by 100~Gbps line-rate with typical Internet traffic.
\maestro further outperforms modern hardware-based transactional memory mechanisms, even for challenging parallel-unfriendly workloads.
\end{abstract}

%% file: sections/1-introduction.tex
\section{Introduction}

With the transition of Network Functions (or NFs) from custom, fixed-function devices to software running on commodity hardware came a well known performance challenge. As line-rates kept increasing, the networking community kept proposing new tools, techniques, and architectural enhancements to overcome individual bottlenecks.
User-mode frameworks, like DPDK~\cite{Intel2010}, bypass the kernel, avoiding costly context switches; DDIO~\cite{DDIO} places incoming packets directly in the CPU cache as they arrive; and NICs implement Receive Side Scaling (RSS)~\cite{rss} to consistently distribute traffic across multiple CPU cores using a configurable hash-function. Despite this wealth of tools, the challenge of developing performant software at these time scales is considerable, typically requiring parallelization~\cite{farshin2021packetmill} and, with it, a deep knowledge of low-level architectural details such as cache-friendly allocation, cache-coherence-aware coordination, and a deep understanding of the RSS hashing mechanism.

Although parallelization is paramount to achieving high performance, ensuring equivalence between parallel and sequential implementations is hard~\cite{khalid2016paving,opennf,split-merge,ftmb,de2014beyond}. Thus, we argue that \emph{developers need not shoulder the burden of fine-grained parallelization themselves}. Much like how developers typically do not write entire code-bases in assembly language, allowing a compiler to analyze their code, extract its functionality, and build an assembly implementation that is equivalent in semantics, we argue that the fine-scaled parallelization of NFs should follow a similar approach. Developers should implement sequential versions of their NFs, benefiting from the inherent simplicity of testing, debugging, and updating such systems, and when deploying to production they can “compile” the NF to obtain its parallelized version.

There are two key insights supporting the solution for this challenge. Due to the increasingly pervasive use of NF frameworks amenable to symbolic execution~\cite{ebpf,katran,xdp,cilium,crab,hxdp,xdp-netdev,hda,zaostrovnykh2019verifying,bolt}, the first key insight is that this technique can be used to not only analyze the NF and infer how it maintains state, but also automatically generate modified versions of it. The second key insight is that by knowing how the NF maintains its state, we can configure the RSS mechanism to send packets accessing the same state to the same core, aiming to minimize inter-core coordination in a parallel implementation, thus maximizing performance.

With these key insights in mind, we propose \textbf{\maestro}, a tool that automatically analyzes a software NF and generates a new implementation that distributes the workload across multiple cores while preserving the semantics of the sequential implementation.
This analysis builds a comprehensive symbolic model of how the NF stores and accesses state, and how that state is structured around flows. Flows (also called \emph{flowspace}~\cite{khalid2016paving} and \emph{scope}~\cite{de2014beyond} by prior work) describe related packets---identified through packet header fields---that the NF logically tracks as an isolated unit.
A firewall, for example, often tracks TCP/UDP flows, identified by the packet 5-tuple (source and destination IPs and ports and the IP protocol number), whereas a traffic monitor may identify flows by destination IP alone.
As NFs typically store state on a per-flow basis \cite{khalid2016paving,mp5}, \maestro learns how flows are defined in the NF by extracting the constraints that define how packets access state.
We then use a solver to find an RSS configuration that distributes traffic across multiple CPU cores, in such a way as to minimize costly inter-core coordination.
Our tool then automatically generates a new implementation of the NF that parallelizes its operation accordingly.

When possible, \maestro generates an implementation based on a \emph{shared-nothing architecture}, wherein RSS is configured to forward packets of the same flow to the same CPU core, completely eliminating any inter-core coordination.
When the NF is not compatible with such a model, \maestro can still generate a parallel implementation where cores share state but accesses to that state are coordinated by a read-write locking mechanism that, while not as performant as a shared-nothing architecture, can still perform well under typical (Zipfian) Internet traffic.

\maestro draws inspiration from prior work in NF analysis~\cite{khalid2016paving} and verification~\cite{Zaostrovnykh2017,zaostrovnykh2019verifying}, as well as the wisdom of a wide body of research on NF performance~\cite{dobrescu2012toward,pedrosa2018automated,bolt,farshin2021packetmill}.
We also use the lessons learned by many before us that address the challenges of \emph{manually} parallelizing NFs, including NUMA considerations~\cite{Emmerich2018}, configuring RSS for symmetric flow handling~\cite{Woo2012}, and rebalancing load with skew~\cite{barbette2019rss++}.

\maestro handles DPDK NFs which store state using the Vigor API~\cite{zaostrovnykh2019verifying}. For these NFs to be amenable to ESE, they are implemented under some constraints, which we describe in~\cref{section:limitations}. These limitations, however, pertain only to NFs given as input to \maestro, and not to the generated parallel solutions.

We evaluate the performance of \maestro by parallelizing 8 DPDK NFs.
Our experimental evaluation shows that NFs that can be parallelized using the shared-nothing architecture scale linearly with the number of cores used until bottlenecked by PCIe when using small packets or by 100~Gbps line-rate with typical Internet traffic~\cite{benson2010network}.
The remaining NFs that require read-write locks to maintain their semantics vary their performance with the workload.
High-churn traffic--where most packets establish a new flow--requires more writing to shared state, degrading performance.
Fortunately, the majority of packets in typical Internet traffic belong to a minority of flows~\cite{benson2010network}, requiring less state writing and allowing more concurrency.
Under this read-heavy traffic, \maestro's lock-based parallel NFs perform comparably to a shared-nothing model.
Notably, when Maestro had to resort to locking, equivalent versions of the NFs that use hardware transactional memory~\cite{larus2007transactional} (TM) to preserve semantics (via the Restricted Transactional Memory interface~\cite{rtm}) were unable to outperform our optimized locks, as we show in \cref{subsection:benchmarks}. We also show that NFs automatically parallelized by \maestro rival in performance with ones manually parallelized using VPP~\cite{barach2018high}.

In \cref{section:motivation}, we describe the inherent challenge of parallelizing NFs, to better motivate our work.
We subsequently present the main contributions of our work, describing the \maestro architecture in \cref{section:architecture} and several key optimizations in \cref{section:implementation}.
In \cref{section:limitations} we discuss \maestro's inherent limitations.
In \cref{section:evaluation}, we evaluate \maestro and the performance of the parallel NFs it generates.
Finally, we describe related work in \cref{section:rw} and conclude with final thoughts in \cref{section:conclusions}.

%% file: sections/2-motivation.tex
\section{Why Parallelization is Hard}
\label{section:motivation}

\begin{figure*}[t]
    \centering
    \includegraphics[width=.8\textwidth]{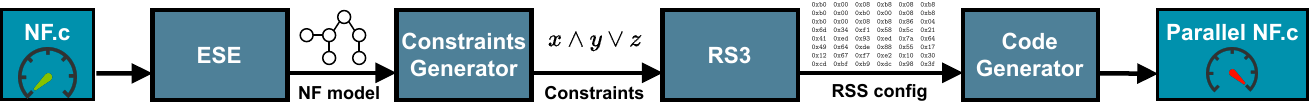}
    \caption{\maestro's architecture}
    \vspace{-1em}
    \label{fig:arch}
\end{figure*}

Ideally, one would parallelize an NF by spinning up individual instances per core, each running independently, and using the NIC to evenly distributing traffic among them. NFs, however, typically store state that persists across packets. Sharing this state among cores requires coordinating access to it, but minimizing this coordination is crucial to achieving high performance. Parallel implementations that require no state sharing among their instances (and therefore no synchronization) are called \emph{shared-nothing}. Implementing a shared-nothing implementation of a stateful NF requires carefully configuring the NIC to distribute traffic to each core in a way that aligns with how state is structured in the NF. With such a mechanism, state is \textit{sharded} across cores and packets accessing the same state always find themselves on the same core.

The NIC can perform this traffic distribution in hardware using the Receive-Side Scaling (RSS) mechanism~\cite{rss}.
This mechanism hashes packet headers using a user-defined set of fields and a hash key.
The computed hash is subsequently used to direct traffic to different queues which can deliver the packets to different cores.
To send, for example, packets of the same TCP flow to the same core, one would configure RSS to hash the source and destination IP addresses, and TCP/UDP ports, and the IP protocol number (\textit{i.e.} the 5-tuple), ensuring that any two packets with the same 5-tuple will have the same hash and will end up on the same core.

This leads us to the traditional method for building parallel shared-nothing NFs: first, developers shard state in the NF, building a full understanding of how state is accessed under all circumstances.
They then use this sharding solution to construct an RSS configuration that distributes traffic accordingly.
This approach, however, poses three big challenges:

\textbf{1. Finding the right sharding solutions is hard.}
Though some NFs simply shard on the 5-tuple, many others require a more careful approach.
One common use case involves symmetrical access to state based on the 5-tuple so that incoming traffic---that has the source and destination swapped---access the same state as outgoing traffic~\cite{Woo2012}.
Other NFs require a more coarse-grained partitioning: some policers and traffic monitors only use the destination addresses to index state, connection limiters may only use source addresses, and network address translators (NATs) will typically shard on the WAN's server address and port (as all the other addresses and ports are translated). Simply sharding on the 5-tuple here would require expensive coordination (\textit{e.g.} locks), as cores are unable to act independently.

Arriving at sharding solutions is harder than generically using locks each time state is accessed.
The developer needs intricate knowledge of the NF's semantics and internals, particularly around how state is kept and manipulated.
This thought process must not only take place upon initial implementation, but also as the NF code evolves over time. Augmenting a firewall with a connection limiter feature renders the previously configured 5-tuple sharding obsolete, requiring a complete rethink of how it should be sharded.

\textbf{2. Finding the right RSS configuration is hard.} Even if we take the sharding solution for granted, configuring RSS accordingly is difficult.
For trivial cases, this is just a matter of selecting the right fields to hash but more complex scenarios can require carefully crafting the RSS key.
Such an approach was used in \cite{Woo2012} to handle symmetrical TCP/UDP flows, but manually tracking the sharding constraints and finding internal symmetries in the hash key that pair with those constraints quickly becomes unmanageable.
For NFs with other sharding requirements, the problem becomes even harder. Not all sets of fields are supported by NICs~\cite{x710,e810}, requiring a specific RSS key that cancels out some bits to circumvents this limitation.
One might even require symmetry between different interfaces (when incoming and outgoing traffic use different NICs), which requires a separate but interrelated configuration and key for each NIC. More complex NFs can shard state in ways that do not neatly fit into any common case, requiring a custom formulation which, as before, may need to be completely rethought from scratch should the NF change over time.
Some cases are outright infeasible, due to inherent NIC limitations, at which point a well-placed warning could help guide developers towards better solutions.

\textbf{3. Writing performant parallel code is hard.} Even if a developer correctly shards the NF and properly configures RSS to achieve a valid shared-nothing solution, they can still be leaving performance on the table. Though shared-nothing goes a long way towards ensuring good performance, many more minute details play a further role in parallel code.
Packet buffers and state must now be cache-aligned to avoid false cache-line sharing. Memory allocation must be NUMA-aware to avoid slower remote accesses across the QPI bus. Even exogenous factors like traffic skew must now be considered~\cite{barbette2019rss++} to fully realize the potential of a parallel implementation.

Getting any of these issues wrong can stand in the way of performance, correctness, or both, but are ultimately amenable to automation. Our tool---\maestro---tackles the first challenge by analyzing how the NF keeps its state and finding the constraints that packets that need to be sent to the same core must satisfy. It further tackles the second challenge by formulating an SMT problem and using a solver to find the right RSS keys that satisfy the sharding requirements. Finally, \maestro addresses the third challenge by automatically generating a parallel implementation that is semantically equivalent to its sequential counterpart. The generated code fully handles NIC initialization and RSS configuration, cache-alignment, load-balancing, and NUMA considerations.
Even when a shared-nothing approach is not possible, \maestro can still help by generating an optimized lock-based parallel implementation that uses carefully crafted read-write locks to minimize inter-core coordination with typical Internet power-law traffic.

%% file: sections/3-architecture.tex
\section{\maestro Architecture}
\label{section:architecture}

\maestro uses symbolic analysis to extract information on how the NF maintains state, and with it infer possible dependencies between parallel instances. This analysis is crucial to achieve synchronization-free parallelization that shards state by carefully splitting traffic among cores. How this careful orchestration of packets can be used to avoid synchronization among parallel instances is better explained via an example.

\subsection{Parallelizing a firewall}
\label{subsection:par_firewall}

Consider a firewall NF connecting a LAN and a WAN that only forwards packets from the WAN that correspond to flows started in the LAN. To keep track of ongoing flows, it stores flow information in a map. Packets from the WAN lookup flow information symmetrically relative to packets from the LAN, naturally swapping source and destination fields.

Note that not all packets need access to all entries in the map: only the ones belonging to the packet's flow. As such, in a parallel execution, making sure that \emph{packets of the same flow are sent to the same core}, conjoined with the fact that packets of the same core are processed sequentially, allows us to parallelize this firewall without any synchronization between its instances---a \emph{shared-nothing} architecture.

This orchestration of packets from the same flow to the same core requires a specific RSS configuration. Not only must we send LAN packets of the same flow to the same core, but also their (symmetric) WAN responses. A configuration partially fulfilling these requirements was already found by Woo and Park~\cite{Woo2012}{\protect\footnote{Woo and Park's solution considers only a single RSS configuration, whereas our firewall deals with two ports (LAN and WAN), each requiring independent configurations. Although their findings are transposable to this scenario, it still requires expertise from the developers.}}. By adapting their configuration to the firewalls' needs, we ensure that every packet that needs access to the same memory region is sent to the same core.

\subsection{Generalizing NF parallelization}
\label{subsection:generalizing}

The above parallelization process is well tailored for our firewall, but different NFs keep state in different ways, and thus require different sharding solutions. Moreover, when access to specific state precludes flow-sharding, synchronization is necessary to maintain semantics.

\maestro deals with this parallelization process automatically by using the architecture shown in~{\cref{fig:arch}}. \maestro starts by analyzing the NF using Exhaustive Symbolic Execution (ESE)~\cite{zaostrovnykh2019verifying,bolt,Cadar2008} to retrieve a sound and complete model of its behavior. Then, it hands the model over to a three stage pipeline: (1) the Constraints Generator, which uses this model to analyze how the NF keeps its state and arrive at a sharding solution; then (2) the RSS configuration generator stage---for which we built a library called RS3---that uses a solver to find an RSS configuration that steers packets following the sharding rules found by the previous stage to the same core; and finally (3) the Code Generator, that generates a parallel implementation that configures the RSS accordingly and adds additional synchronization mechanisms if needed.

\subsection{Extracting the NF's model}

\maestro uses ESE to extract the complete NF's model. This allows us to not only analyze how the NF maintains its state, but also generate modified versions of its implementation.

The extracted model is an execution tree containing all the possible code execution paths a packet can trigger. Each node on this graph is either conditional (representing a branch condition), a stateful operation (representing a call to a stateful data structure, \eg a map or a vector), or packet operation (\eg forwarding, dropping, etc.). Both the packet and stateful data are traced as symbols, and every node contains a list of constraints on these symbols that can be given to a solver to query their possible values under any code path.

\subsection{Finding the sharding solution}
\label{subsection:sharding}

The NF model is passed to the Constraints Generator, which is tasked with finding a sharding solution that allows shared-nothing parallelization.
The idea is to find the constraints that hold true between packets that access the same state, \ie packets that must be processed on the same core.
This is intrinsically tied to how the NF maintains state.
For example, in a map for two operations to access the same state they must use the same \emph{key}.
By symbolically tracking how such keys are derived from packets, we reason about the constraints on packets that access common state.

\subsubsection{Building a stateful report} The Constraints Generator starts by analyzing the NF's model and builds a stateful report (SR) of all the performed stateful operations. Each SR entry specifies the operation's name (\eg \texttt{map\_put}), object instance, and other relevant arguments (\eg the key used), and all the possible constraints on both the received packet and other stateful data when the operation was performed (\eg \texttt{map\_put} was called when a UDP packet arrived from interface 0).

\subsubsection{Filtering entries} After building the SR, the Constraints Generator removes all entries related to read-only objects (\eg routing tables that are filled on start-up and never updated).
Such read-only accesses to shared state do not require coordination among cores and need not be reasoned about.
Should all accesses be read-only, the SR will be left empty and \maestro asks the Code Generator to generate a parallel implementation that uses RSS with the sole purpose of load-balancing traffic among cores (we explain the RSS mechanism in \cref{subsec:keygen}).

\subsubsection{Analyzing the entries} The use of any data structure can potentially preclude a shared-nothing approach, and therefore we need to infer the conditions under which it is safe to perform stateful operations concurrently for each of them (or if no such conditions exist). We present the analysis for one of the most predominant data structures: the map~\cite{khalid2016paving,zaostrovnykh2019verifying,ebpf-maps,khalid2016paving}.

\begin{figure}[t]
    \centering
    \includegraphics[width=\linewidth]{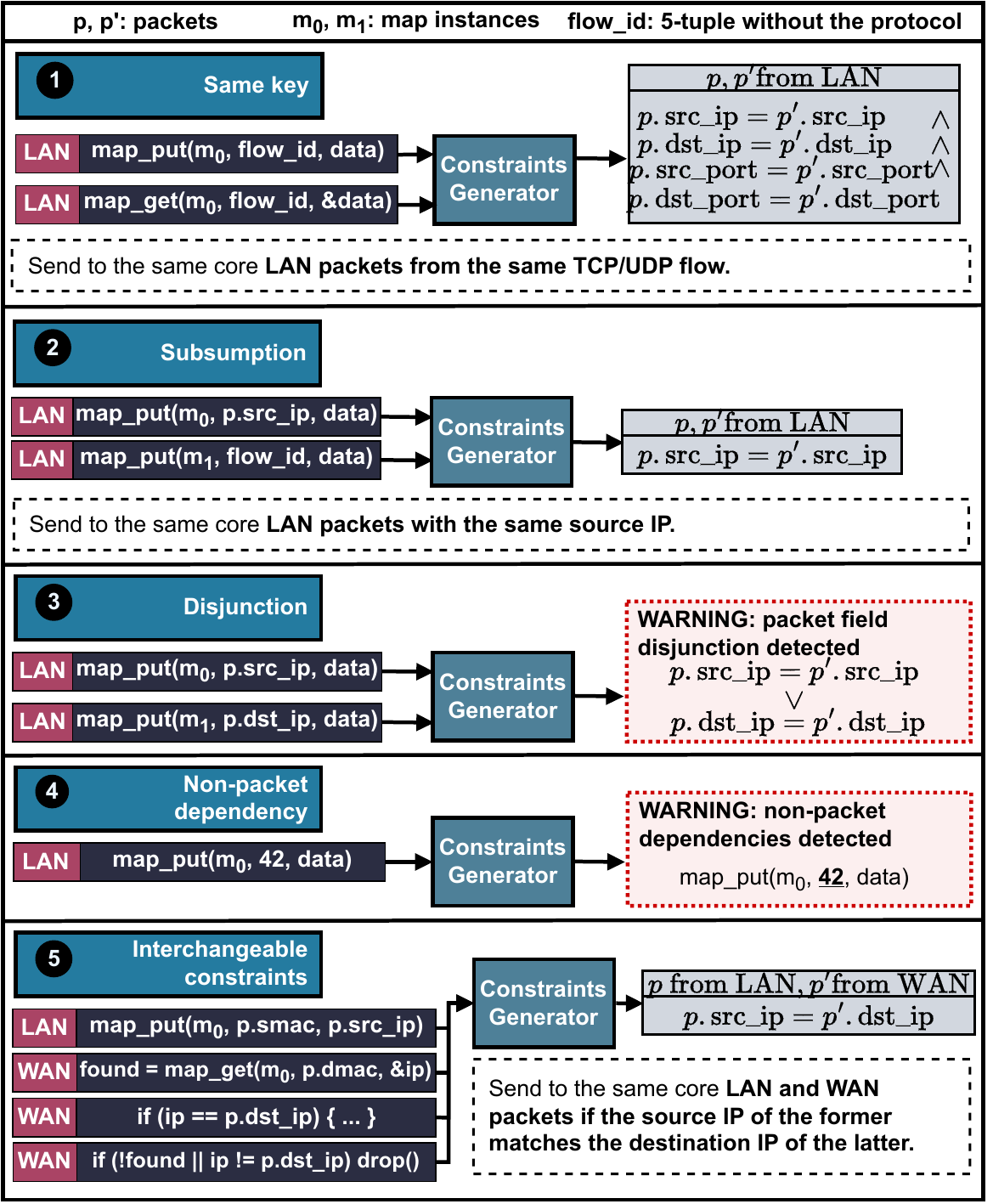}
    \caption{Example outputs of the Constraints Generator.}
    \label{fig:constraints_generator}
    \vspace{-2em}
\end{figure}

The map stores data indexed by a key. This data can be accessed via the function \texttt{map\_get}, and modified with \texttt{map\_put}. Two map calls access the same memory region if and only if they are given the same key. For a shared-nothing approach, packets that trigger map calls to the same instance using the same key need to be steered to the same core. This alone is, however, insufficient: we need to not only take into consideration any RSS limitations, but also reason about the use of multiple different map instances (or other data structures), each independently tied to the previous requirement. With this in mind, we designed a set of rules to guide \maestro towards finding correct shared-nothing sharding solutions:

\begin{enumerate}[label=\textbf{R\arabic*},ref=R\arabic*,noitemsep]
    \item \label{cf:eq_keys} \textit{Key equality.} The most obvious case is when two packets access the same map instance using the same key. In this case, the Constraints Generator builds the constraint from the formulas for the keys (\circled{1} in \cref{fig:constraints_generator}).
    
	\item \label{cf:subsumption} \textit{Subsumption.} If a map instance is accessed using a subset of the packet fields used to access a second instance, then the subset takes precedence over its larger counterpart.
    That is, the coarser-grained requirement wins over the finer-grained one.
    This is exemplified in scenario \circled{2} in \cref{fig:constraints_generator}: sending packets with the same source address to the same core will also guarantee that packets with the same 5-tuple are also sent to the same core. More generally, we can always use a subset of the required packet fields. As we will see later, this rule can act further in concert with others to resolve incompatibilities.

	\item \label{cf:disjoint} \textit{Disjoint dependencies.} Accesses using disjoint sets of packet fields are problematic. An NF that keeps a pair of independent counters, one for source addresses and another for destination addresses, requires packets with the same source address \emph{or} the same destination address to be sent to the same core. Due to limitations in the RSS mechanism, this is not possible: configuring it with both the source and destination fields will guarantee that packets with the same source \emph{and} destination will be sent to the same core. \maestro warns the user and provides the fundamental reason why the shared-nothing approach cannot be applied (\circled{3} in \cref{fig:constraints_generator}).

	\item \label{cf:incompatible} \textit{Incompatible dependencies.} RSS uses packet fields to steer packets to cores. This means that using keys containing (1) incompatible RSS packet fields or (2) no packet fields at all will completely block our attempt at correctly steering packets to cores. This is the case, for example, of NFs which index data with constant keys, as exemplified in case \circled{4} of \cref{fig:constraints_generator}.
    Again, in this case, \maestro provides feedback to the user as to why the shared-nothing approach is unfeasible\protect\footnote{\maestro behaves in a similar manner when finding global counters updated by every packet, as it bars it from implementing a shared-nothing parallel solution.}.

	\item \label{cf:interchangeable} \textit{Interchangeable constraints.} We define a pair of constraints as \emph{interchangeable} if they trigger the same NF behavior. This allows us to completely replace constraints matching rules \ref{cf:disjoint} or \ref{cf:incompatible} with others that, if interchangeable, do not prohibit shared-nothing parallelization.

    Example \circled{5} of \cref{fig:constraints_generator} showcases this scenario. In this example, the packet is dropped when we fail to find the MAC address entry on the map, or whenever the incoming IP does not match with the stored address. Although the NF stores source addresses using an RSS-incompatible dependency on our NIC~\cite{e810} (source MAC), the Constraints Generator finds that the NF's behavior is \emph{exactly the same} whether we shard on the MAC address or the destination address. In this case, these constraints are interchangeable, which allows \maestro to shard on either of them. Because the former uses an incompatible RSS field, the Constraints Generator opts for using the latter one for sharding.

    By sharding with the IP address, changing solely this field can cause the packet to be sent to a different core. Although it may still find a matching entry of its MAC address on the map, it will find a different IP address stored on that same entry, and hence the packet will be dropped. Both not finding the MAC entry and mismatching the IP value result in the same behavior from the NF.
\end{enumerate}
\vspace{-2em}

These rules allow \maestro to correctly find sharding solutions for a wide range of NFs (as we show in \cref{section:evaluation}).
Note that only \ref{cf:eq_keys} is specific to data structures that use a key to index state (\eg maps, vectors, sketches). \ref{cf:subsumption}, \ref{cf:disjoint}, \ref{cf:incompatible}, and \ref{cf:interchangeable} are otherwise data structure agnostic, and \maestro employs them to all entries, regardless of their specific data structure.
Though much of this analysis focuses on maps, it can be used as building blocks for others.
Moreover, we need only reason about these details \emph{once} per data-structure (or, at most, each time a breaking change is made).
Once data-structure developers encode such properties into \maestro, NF developers can freely use these stateful data structures to build their NFs.
\cref{table:stateful-constructors} shows the stateful constructors currently supported by \maestro.

Even when \maestro fails to find a shared-nothing solution, it still provides the developer the fundamental reason why (\eg constant keys or non-packet dependencies). When met with this result, the developer is faced with a decision: either use this feedback to tweak the NF implementation so that it becomes amenable to shared-nothing parallelism, or request a lock-based implementation from \maestro.

\begin{table}[t]
\centering
\begin{tabular}{|l|l|}
\hline
\textbf{Name} & \textbf{Description} \\
\hline\hline
map & Stores integers indexed by arbitrary data. \\ \hline
vector & Stores arbitrary data indexed by integers. \\ \hline
dchain & Time-aware integer allocator. \\ \hline
sketch & Count-min sketch~\cite{count-min-sketch}. \\ \hline
\end{tabular}
\caption{Stateful constructors currently supported by \maestro.}
\label{table:stateful-constructors}
\vspace{-2em}
\end{table}

\subsubsection{Generating the constraints}
The next step in the \maestro pipeline is to generate the actual constraints, \ie, the conditions that, if satisfied by a pair of packets, dictate that they must be sent to the same core.
Towards this end, \maestro iterates over each pair of report entries of the same state instances, creating SMT formulas stating that both keys must be equal, and joining them all together with logical \emph{OR}s.

Finally, we note that RSS must be independently configured on each interface. As such, the constraints generated by \maestro are interface-specific, reasoning about pairs of packets which may arrive from separate interfaces. Case \circled{5} from \cref{fig:constraints_generator} exemplifies this. It requires LAN packets to be sent to the same core as packets from the WAN if the source address of the former equals the destination address of the latter.

\begin{figure}[t]
    \centering
    \includegraphics[width=\linewidth]{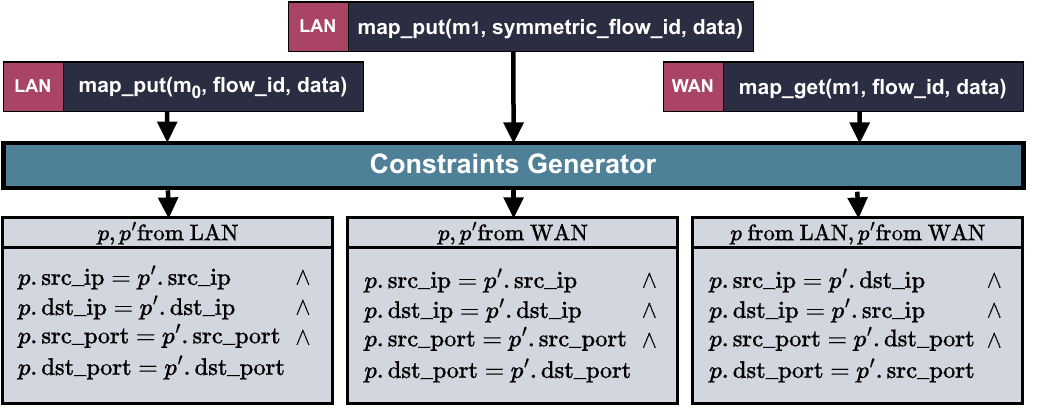}
    \caption{From the firewall's SR to its sharding constraints.}
    \label{fig:fw-pipeline}
    \vspace{-2em}
\end{figure}

\cref{fig:fw-pipeline} shows the constraints found by the Constraint Generator when analyzing our firewall example. It finds that LAN packets with the same addresses and ports must be sent to the same core, and similarly for WAN packets. It also finds that WAN and LAN packets must be sent to the same core if they have the same, but swapped, sources and destinations.

\subsection{Finding the right RSS configuration}
\label{subsec:keygen}

The previous stage tackled the challenge of finding a shared-nothing sharding solution, producing constraints between packets that when true require the packets to be processed on the same core. We now focus on materializing this sharding solution by automatically finding RSS configurations that satisfy these constraints.

RSS is a hardware mechanism in the NIC that steers packets to core-specific queues. Once configured with an RSS key and a set of packet fields, it extracts from incoming packets the values of those fields and feeds them to a toeplitz-based hash-function~\cite{rss-hashes}. This function, depicted in \cref{fig:toeplitz}, works by continuously left rotating the key $k$ while iterating through the selected packet fields bits $d$. The running 32-bit hash value is \emph{XOR}'ed with the current $32$ least significant bits of the key whenever the current bit $d_i$ is 1. The resulting hash is used to index an indirection table containing queue identifiers, and the packet is inserted in the corresponding queue.

Two packets with the same hash will be sent to the same core. Given the configurability of the RSS hashing function, we use it to ensure that packets that need to be processed on the same core will have the same hash.
For simple constraints we can arrive at a satisfying RSS configuration solely by correctly choosing the packet field set (\eg, hashing only source and destination IPs and ports when requiring TCP packets with the same 5-tuple to be sent to the same core).
However, what if (1) the NF requires a subset of packet fields that can only be used as a group in the RSS mechanism (\eg, a traffic monitor that shards solely the destination IP), (2) it requires complex constraints between packets (\eg, a Hierarchical Heavy Hitter sharding on multiple subnets of the source IP and/or source ports), or (3) there are constraints between packets arriving in different interfaces (which is the case for many NFs requiring both LAN and WAN interfaces, as in NATs, Firewalls, Connection Limiters, \etc)?

To address these scenarios in a generalized way, we built \librs, a C library capable of taking constraints as inputs and outputting RSS configurations that satisfy them. It uses the Z3 solver~\cite{de2008z3} to find suitable configurations by encoding the problem in a logical format. \maestro uses \librs to generate RSS configurations that satisfy the constraints given by the Constraints Generator module.

\subsubsection{Building the statement}
The query given to the solver needs to encode the following problem: \emph{given set of constraints, find RSS keys that generate the same hash for every pair of packets that satisfy them}. To build this statement, we need to encode both the hash function and the constraints into an SMT format.

Let $k$ be a 52~byte\protect\footnote{Value for the Intel E810 100G NIC~\cite{e810}, but trivially adjustable in \librs.} RSS key, $d$ and $d'$ hash inputs for each of the packets (whose sizes depend on the extracted packet fields, \eg 12~bytes for source and destination IPs and ports), and $h(k,d)$ the 32~bit hash. Also, let $|k| \ge |d| + |h|$, $H(k, k', d, d')$ be true iff $h(k,d) = h(k',d')$, and $C(d,d')$ be the constraint between $d$ and $d'$ provided by the constraint generator.

\begin{figure}[t]
    \centering
    \includegraphics[width=\linewidth]{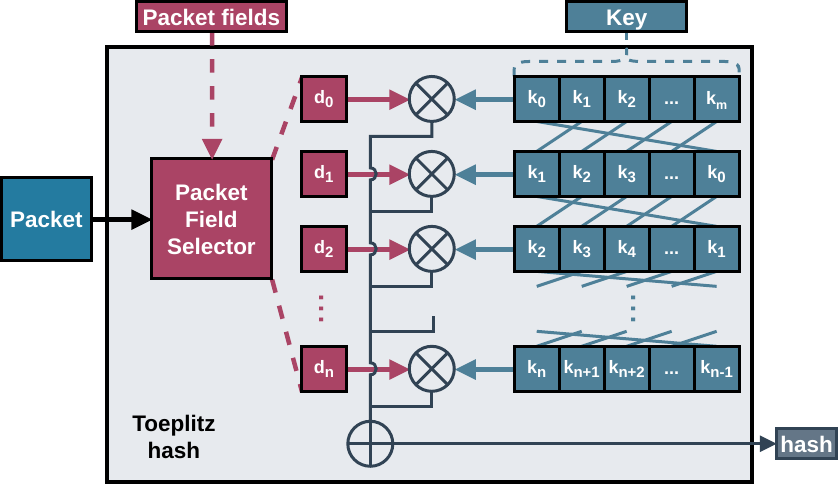}
    \caption{Toeplitz-based hash function.}
    \label{fig:toeplitz}
    \vspace{-2em}
\end{figure}

\noindent
\textbf{Hash function.} As shown in \cref{fig:toeplitz}, $H(k, k', d, d')$ can be represented as:

\vspace{-1.5em}
\begin{align}
\label{eq:hashes_equality}
\bigwedge_{b = 0}^{|h|-1}
\left[
\bigoplus_{x = 0}^{|d|} ( d[x] \land k[x + b] ) = \bigoplus_{y = 0}^{|d'|} ( d'[y] \land k'[y + b] )
\right]
\end{align}

Note that although the size of the key is lower bounded, it should not have any influence on the feasibility of finding a suitable hash configuration. Only a subset of its bits are used on the hash function, and therefore constrained by our requirements, with the other bits being free to take any value.

\noindent
\textbf{Base statement.} Initially, let us encode the following query: \emph{find a single key $k$ such that, given any two hash inputs $d$ and $d'$ that obey the constraints C, their corresponding hashes will always be equal.} That is:

\vspace{-1.5em}
\begin{equation}
\label{eq:statement_non_trivial}
\forall_{d, d'} \dsepA
k \ne 0 \land \left[
(C(d, d') \land d \neq d') \rightarrow H(k, k, d, d')
\right]
\end{equation}

Having the key be 0 would always output 0 valued hashes, steering all packets to a single core, so we prevent the key from taking that value.

\noindent\textbf{Compatibility with multiple keys.} Each interface can have its RSS mechanism individually configured.
With that in mind, let $C_{ij}(d,d')$ be the constraint between a pair of packets coming from ports $i$ and $j$, configured with the keys $k_i$ and $k_j$ respectively.
Note that $C_{ij} = C_{ji}$, therefore it is enough to consider, for example, all the constraints $C_{ij: \{ j \le i \}}$.
For \cref{eq:statement_non_trivial} to be multi-key aware, we simply conjunct the constraints across all $i$ and $j$, allowing the solver to manage each key combination problem as a specific statement that must be true. That is, for $n$ ports:

\vspace{-1.5em}
\begin{equation}
\label{eq:statement_multikeys}
\forall_{d, d'} \dsepA \bigwedge_{i = 1}^n \bigwedge_{j = 1}^i \left[ ( C_{ij}(d, d') \land d \neq d') \rightarrow H(k_i, k_j, d, d') \right]
\end{equation}
\vspace{-1em}

\noindent\textbf{Compatibility with varying sets of RSS packet fields.}
Just as different ports may need distinct RSS keys, we may also need to configure RSS to use different sets of packet fields depending on the interface. One way to address this would be to consider hash inputs $d_0, ..., d_{n-1}$ for $n$ interfaces. This, however, highly increases the complexity of the query, making it harder for the solver to find a solution\protect\footnote{For $n$ interfaces, and thus considering $d_0,d_0',...,d_{n-1},d_{n-1}'$, with 96~bit hash inputs we would have to deal with $2 \times 96 \times n$ free bits.}.
Another way to look at it would be to extend the hash inputs to include the union of both field-sets and to deal with any unused bits.
To make the statement in \cref{eq:statement_multikeys} consider constraints between packets arriving at different ports with different RSS packet field options, we again add more clauses to our large conjunction, now considering all relevant RSS field sets, all while extracting for each one the required least significant bits of $d$ and $d'$ accordingly.

When given the constraints of our firewall, \librs outputs two RSS keys, one for each NIC interface. The symmetry between the keys resembles the findings in \cite{Woo2012}, but generalized to two interfaces, rather than just one.

\subsection{Code Generator}
\vspace{-0.5em}

This stage takes the generated RSS configuration, as well as the NF's model, and outputs a parallel implementation of the original NF. Because the model is a sound and complete representation of the original NF, it can be used to generate an implementation identical in functionality to the original one. More importantly, it can be modified to employ shared-nothing parallelism by (1) configuring RSS, (2) allocating the state independently for each core, (3) making sure that each stateful call uses the data structures' instances of that particular core, and (4) launching the NF in multiple cores. \cref{appendix:code} contains adapted code excerpts from our Firewall example, showing both the sequential implementation used as input to \maestro and the final generated parallel shared-nothing implementation.

\subsubsection{Parallel implementation with locking mechanisms}
When \maestro rules out a shared-nothing solution, it can fall back to generating parallel implementations that use locking mechanisms.
In this scenario, it configures RSS with both a random key and all the available RSS-compatible packet fields, as now all cores share the same state.

\maestro also needs to carefully coordinate access to shared data using read/write locks.
As such, we distinguish read-packets from write-packets: the former trigger only stateful read operations, and the latter trigger at least one write.
To efficiently handle this scenario, we created a custom, highly optimized read/write lock implementation that entirely avoids cache-line sharing when acquiring read locks.
We do this with a series of per-core, cache-aligned, atomic spin-locks that indicate whether the core has permission to proceed.
Acquiring a read lock requires just locking the current core's lock. To perform a write, however, one must lock all core-specific locks (in order, to avoid deadlocks).
With this in place, we speculatively process all packets as read-only until they attempt to perform a write operation, at which point we stop processing, release the local lock, acquire all core-specific locks, and restart processing the packet from the beginning.

The performance toll is minimized when an NF is subjected to read-heavy workloads (see \cref{subsection:benchmarks}), as read-only packets need only acquire a core-specific cache-aligned lock, and have no need to atomically write to any shared variable, or write to shared data. As all write-packets start out as read-packets before backtracking, starvation is not an issue.

%% file: sections/4-implementation.tex
\section{Implementation challenges}
\label{section:implementation}

\subsubsection{Finding good RSS keys}
The first set of keys found by the solver is often not ideal.
If, for example, the solver finds a key with all but the first bit set to zero, the hash, though semantically valid, will only ever be $\mathtt{0x0}$ or $\mathtt{0x80000000}$. This leads to packets being sent to only two cores.

The solution employed by \librs involves setting the value 1 to as many bits as possible in the keys, so long as they still satisfy the given statement. This is known as a Partial MAXSAT problem \cite{cha1997local}.
We give the solver a statement that its corresponding solutions should always satisfy---\cref{eq:statement_multikeys}, hard constraints---and also a set of clauses that they should try to satisfy---soft constraints. The soft constraints correspond to a chain of logical \textit{ANDs} setting each key bit to 1.
There is no need for maximizing the number of satisfied soft constraints.
Most of the times, a randomly selected set of bits with the value 1 is enough to avoid corner case problems like the one mentioned above.
As such, \maestro uses a slightly modified version of the diagnosis-based approach introduced by Fu and Malik \cite{fu2006solving}.
It begins by seeding the key with random bits.
Then, if the combined hard and soft constraints are not satisfiable, we get the UNSAT core from the solver and randomly discard a subset of these soft constraints, repeating as necessary until either a key is found or no further soft constraints are left, indicating that no such key exists.
Due to the randomized nature of this algorithm, we use multiple parallel solvers to independently find keys until one is found with an acceptable workload distribution.

\subsubsection{NUMA considerations}
In a NUMA environment, each possible combination of NIC, memory, and CPU pinning influences throughput. Our machines (see \cref{section:evaluation}) have 100~Gbps NICs with 2 interfaces, thus both interfaces are pinned to the same NUMA node.
Under these circumstances, pinning the packet buffers to the same NUMA node as the NIC is optimal~\cite{Emmerich2018}.

Another important consideration is that the dominant contention factor in parallel packet processing applications is the cache, specifically for Intel Data Direct I/O (DDIO) resources~\cite{dobrescu2012toward,manousis2020contention}. Using DDIO, the packets coming from the NIC are directly placed in the last level cache (LLC) of the NUMA node.
Contention happens when the number of concurrent packets exceeds the available reserved space for I/O in the LLC, at which point packets evict each other and performance suffers.
\maestro allocates packet buffers close to the NIC, but keeps state local to each core's NUMA node. Deciding where to run each thread is, however, a deployment challenge, not an implementation one, and therefore out of scope for \maestro. Nevertheless, our experience has taught us a simple rule of thumb: if the LLC is large enough to hold all packet buffers at line-rate, then we should pin both the CPU and memory to the same NUMA node as the NIC. If, however, the LLC is too small, resulting in contention---as occurs with older processors---then it's better to distribute cores evenly across NUMA nodes, thus increasing the total available LLC.
Though we have seen scenarios where using multiple NUMA nodes was best, in our testbed the LLC proved sufficiently large to justify using a single NUMA node, and all our experiments in this paper follow this guideline.

\subsubsection{Traffic skew}
\label{subsec:skew}
{The expression "mice and elephants" is typically used to describe packet flow distributions on the Internet \cite{benson2010network,guo2001war,lan2006measurement}.
These follow a Zipfian distribution, where a large fraction of packets relate to but a few flows, and the remaining ones share a small slice of traffic.}

\begin{figure}[t]
\centering
	\includegraphics[width=\linewidth]{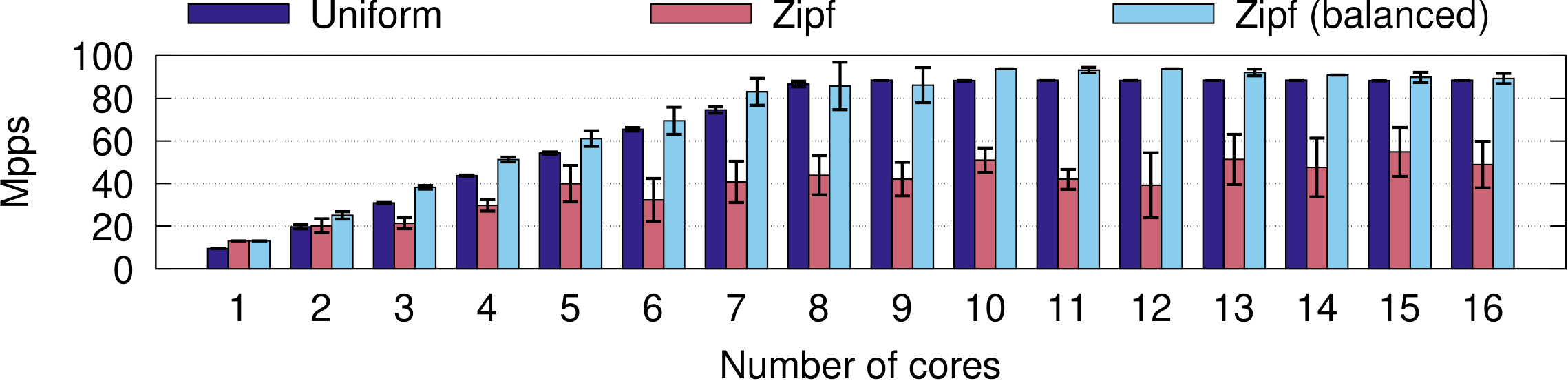}
 	\caption{Shared-nothing firewall under uniform and Zipfian traffic, with and without balanced tables.}
 	\label{fig:traffic}
    \vspace{-2em}
\end{figure}

While traffic with a uniform distribution leads to packets being uniformly distributed to cores, traffic following a Zipfian distribution can overload a subset of cores, causing \emph{skew}. This performance difference is shown in \cref{fig:traffic}, which demonstrates how the parallel firewall throughput varies with the traffic distribution.
The Zipfian traffic was generated with parameters from \cite{pedrosa2018automated}, which were found by analyzing a real-world traffic sample from a University network in \cite{benson2010network}. This generated traffic has 50k packets and 1k flows, 48 of which responsible for 80\% of the traffic.
RSS was configured with five different random keys and the error bars represent the min/max performance.
Performance is influenced by both the RSS key and the indirection table, as more hash collisions cause more packets being sent to the same core.
Under uniform traffic, the indirection table's entries are expected to be equally accessed, and thus uniformily filling it leads to evenly spreading packets across cores.
With Zipfian traffic, however, the higher density of certain flows leads to more accesses to some entries, overloading some cores.
Note that when using a single core we see better performance under Zipfian traffic due to an increased cache hit-rate when accessing state~\cite{pedrosa2018automated}, though the effect is less prominent when more cores are used.

RSS++~\cite{barbette2019rss++} fixes the distribution problem imposed by Zipfian traffic by dynamically adjusting the indirection table according to the traffic. It balances the indirection table by swapping entries associated with overloaded cores for ones associated with underloaded ones. 
It also provides us with mechanisms for state migration across cores which avoid both blocking and packet reordering. We implemented static versions of these mechanisms in \maestro, but their dynamic versions could be used to handle changes in skew over time.

\subsubsection{State sharding}
\label{subsec:sharding}
When applying shared-nothing parallelization, \maestro not only allocates each data structure instance on each core, but further adjusts each data-structure's capacity, keeping approximately constant the total amount of memory used for all cores by reducing the per-core amount.

This raises an interesting question about the semantics of filling up state in a shared-nothing parallel version of an NF, which slightly differs from the sequential or lock-based parallel versions.
As each core now has a reduced capacity, it is possible to exhaust the capacity of one core despite there being spare room in others.
Ultimately, when a core becomes ``full'', it will behave in the same way locally as the sequential NF would globally (\eg by dropping packets from new flows).
As the RSS++ mechanism redistributes flows across cores to counteract traffic skew, this also affects state distribution, making it harder to exhaust any one core.

This state sharding has the desirable side-effect of optimizing the NF's cache utilization.
If each core has a smaller working-set, more of it will fit in the local L1+L2 data caches.
This provides an extra performance advantage to the shared-nothing approach on top of that of parallelization on its own.

\subsubsection{Lock-based rejuvenation}
When following a read-write lock-based parallelization approach, flow rejuvenation can be a challenge.
As simply reading state requires updating the flow entry aging data, a naive implementation would require a write lock for all packets, with dire consequences for performance. \maestro circumvents this issue by implementing an optimized rejuvenation algorithm that operates locally in each core for most cases.
We first modify the data-structures to hold multiple cache-aligned copies of the entry aging data, one per core.
Each core then manages state aging locally for each entry, allowing the age of the entries to deviate from core to core as packets from the same flow arrive at different cores at different times.
When eventually one core believes it should expire an entry, only then does it acquire a write lock.
At this point, the core inspects the aging data for that entry on all cores. If the flow indeed expired on all cores, it is cleared out globally. If, however, another core is found where the entry has not yet expired, the local timestamp is re-synced with the newest one. Ultimately, if packets from the same flow regularly hit all cores, no write-locks are ever needed.

\subsubsection{Implementation}
\maestro uses the KLEE symbolic execution engine, extending it with 14,859 lines of C++ code. We also implemented \librs in 3,964 lines of C code, independently from \maestro\protect\footnote{Our code is openly available at~\cite{maestro-repo}.}.

%% file: sections/5-limitations.tex
\section{Assumptions and limitations}
\label{section:limitations}

\subsubsection{NF limitations}
To allow ESE, NFs must fit within some limitations, much like the ones enumerated in \cite{pix}:
i) there must be a clean separation between stateful and stateless operations, a constraint put in practice by only allowing state to persist within a set of well-defined data structures;
ii) loops must be statically bounded;
and iii) no pointer arithmetic is allowed outside the data-structures. These constraints are already enforced for safety reasons in commonly used packet processing framework like eBPF\footnote{NFs developed in eBPF store their state in kernel-maintained maps~\cite{ebpf-maps}.}~\cite{ebpf}, a widely used framework in both academia and industry\footnote{There is an ongoing effort in adapting \maestro to also accept eBPF NFs as input (an effort already set in motion by PIX~\cite{pix}).}~\cite{katran,xdp,cilium,crab,hxdp,xdp-netdev,hda}.

\subsubsection{RSS limitations}
For \maestro to consider other hash function besides the standard toeplitz-based one, they would have to be formulated as an SMT problem and added to \librs. This requires having their implementation openly disclosed.

In practice, a more limiting factor is packet field selection: shared-nothing approaches can only be applied if state is sharded using RSS-compatible packet fields. DPDK's API~\cite{dpdk-rss-fields} reference includes all possible field combinations that RSS can use (\eg IPv4/IPv6 TCP/UDP flow tuples), but each NIC only implements a subset of them~\cite{e810,x710}.

\subsubsection{Attacking state sharding}
We mentioned earlier that it would be possible to ``fill-up'' a single core with fewer flows in a shared-nothing parallel NF than would otherwise be needed in the sequential or lock-based parallel versions.
This could potentially be used as a DoS attack vector, reducing the cost for an attacker to block new flows from being admitted.
RSS++ flow redistribution addresses this for well-behaved traffic, but an attacker can subvert this by specifically using flows that induce exact RSS hash collisions.
Colliding flows end up on the same entry within the RSS indirection table and thus cannot be split apart.

Though out-of-scope for this paper, \maestro provides some defense from such attacks due to the randomization used to generate RSS keys.
Even assuming the attacker has access to the NF source code and understands how it can be sharded across cores, different random RSS keys that comply with the sharding constraints will still distribute different flows in a different way.
Without access to the actual key generated in \librs, the attacker would have a harder time reverse-engineering a set of co-located flows, mitigating their ability to induce the kind of persistent skew needed in a successful attack.


%% file: sections/6-evaluation.tex
\section{Evaluation}
\label{section:evaluation}

In this section, we evaluate \maestro and the three different types of parallel implementations it can generate: (1) \emph{shared-nothing}, (2) \emph{lock-based}, and (3) parallel solutions using \emph{hardware transactional memory}~\cite{larus2007transactional} via the Intel's Restricted Transactional Memory interface~\cite{rtm}.
We aim to answer four questions:
(i) how long does it take \maestro to parallelize NFs?
(ii) how well does the performance of these parallel implementations scale with the number of cores?
(iii) what are the impacts on performance of the various parallelization strategies that \maestro can use?
and (iv) how do \maestro's automatic parallel implementations fare against highly-optimized manually parallelized versions?

\subsection{Target NFs and Microbenchmarks}
\label{subsection:microbenchmarks}

\begin{figure}[t]
    \centering
    \includegraphics[width=\linewidth]{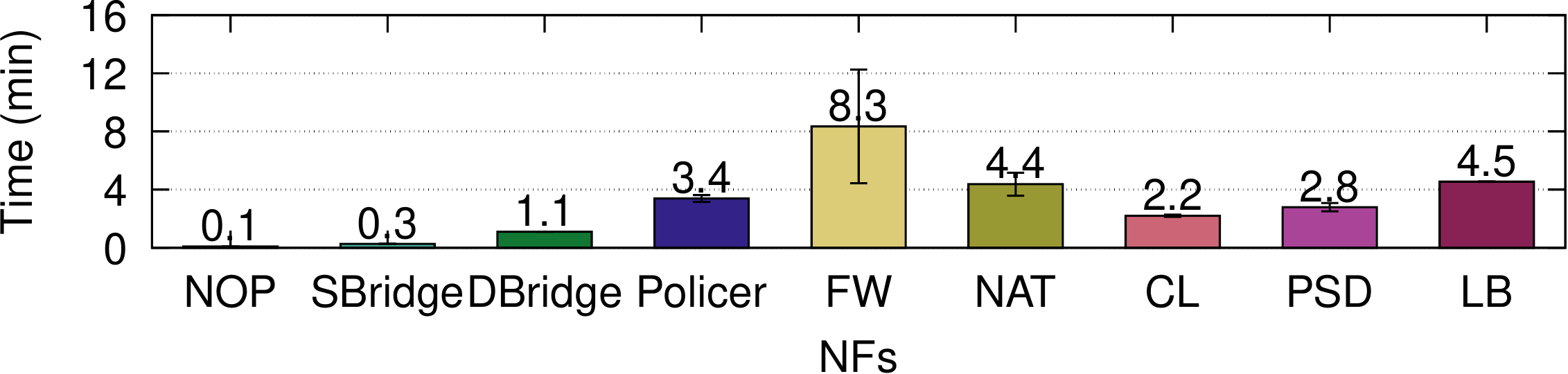}
    \caption{Time (in minutes) to generate parallel implementations for each NF (averaged over 10 runs).}
    \label{fig:microbenchmarks}
    \vspace{-1.5em}
\end{figure}

To evaluate \maestro we analyzed 8 NFs---a simple forwarder (NOP), a policer, a bridge, a firewall (FW), a port scan detector (PSD), a NAT, a load-balancer (LB), and a connection limiter (CL). These are open-source NFs, most are non-trivial in complexity, and all have been used by a body of previous work~\cite{zaostrovnykh2019verifying,bolt,pix}\protect\footnote{As mentioned in \cref{section:limitations}, the requirement that NFs be amenable to ESE can prevent \maestro from analyzing many existing codebases.}.
In this section, we present a brief description of each, and show how \maestro parallelizes them\protect\footnote{Every automatically generated parallel solution can be found on~\cite{maestro-repo}.}.
For each NF, we measured how much time \maestro took to generate a parallel implementation (shared-nothing when possible, lock-based otherwise), summarizing the results in \cref{fig:microbenchmarks}.

\subsubsection{NOP}
This is a simple forwarding no-operation NF, \ie a stateless NF that simply forwards all packets that arrive from one interface to the other.
\maestro finds that this NF has no state, and provides no constraints between packets arriving at the same core.
RSS is thus configured with all available packets fields and a random key on both ports.

\subsubsection{Policer}
This NF aims to limit each user's download rate, identifying users by their IPv4 address.
When \maestro analyzes this NF, it finds that state is indexed by the destination IP address, implying that packets with the same destination address must be sent to the same core.
Because this constraint uses the destination IP address, the chosen RSS packet field options must contain this field.
Although DPDK allows RSS packet field options containing only IP addresses, our NICs do not support this option.
\maestro thus chooses a packet field option that includes IP addresses and TCP/UDP ports.
This increases the complexity of the constraints on the key, increasing the generation time in \cref{fig:microbenchmarks}.

\subsubsection{Bridge}
A bridge associates MAC addresses with interfaces, and redirects packets accordingly.
In a typical MAC learning bridge, the association between source MAC addresses and input interface is learned dynamically.
When analyzing this NF, \maestro detects that state is indexed by a MAC address, which is a field not supported by RSS on our NIC.
As such, \maestro warns the user that it cannot generate a shared-nothing implementation, opting for read/write locks instead.

By modifying the NF to disable dynamic MAC learning, leaving only statically configured MAC-Port bindings, the NF becomes more amenable to parallelization (as all state is read-only), albeit with reduced functionality.
This further illustrates the ability of \maestro to inform developers and help guide the development process by pointing out relevant trade-offs between functionality and performance.
With this in mind, we created two versions of this NF: the standard bridge with dynamic MAC learning (DBridge) and a static one with fixed bindings (SBridge).
When analyzing SBridge, \maestro encounters only read-only data structures, requiring no specific constraints on the RSS configuration.
As with NOP, \maestro generates a random RSS key and uses all the available packet fields on all ports.

\subsubsection{FW}
\label{subsubsection:microbenchmarks:fw}
This is the same firewall we have been using as a running example throughout the paper (\cref{subsection:par_firewall}). It indexes state with typical flow information on the LAN (source and destination addresses and ports), and symmetrically on the WAN. \maestro generates a shared-nothing implementation that shards state by the flow information, sending WAN packets corresponding to symmetric LAN sessions to the same core as these (as shown in \cref{fig:fw-pipeline}).

\subsubsection{PSD}
A Port Scan Detector (PSD) counts how many distinct destination TCP/UDP ports each host (source IP) has touched within a given time frame.
Above a threshold, connections to new ports are blocked, preventing port scans.
\maestro analyzes the PSD and finds that it uses only the source IP to access one map, but also the source IP and destination port to access another. As such, the constraints for accessing the first map subsume those of the second (\ref{cf:subsumption}) and \maestro finds an RSS key that shards based only on source IPs.

\begin{figure}[t]
    \centering
    \includegraphics[width=.7\linewidth]{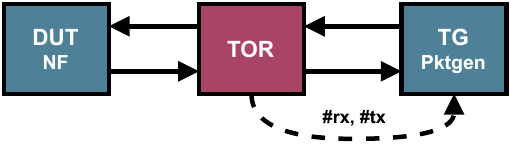}
    \caption{Testbed for our experiments.}
    \vspace{-2em}
    \label{fig:testbed}
\end{figure}

\subsubsection{NAT}
A NAT translates addresses between a LAN and a WAN, allowing multiple clients in the LAN to share a single public IP in the WAN~\cite{rfc3022}.
It keeps track of flows initiated in the LAN, but to aid with translation it associates a unique external port with each flow.
Reply packets from the WAN are checked to see if their address and port match those on record before subsequently translating the destination address and port to match those of the client.

\maestro notices that the NAT associates flows with external ports using a map, fitting case \ref{cf:incompatible} in \cref{subsection:sharding}. However, it also finds an additional constraint fitting case \ref{cf:interchangeable}: packets from the WAN are only translated if they target the hosts that started the session in the first place. This constraint allows for sharding based on the external server's IP address and port.

Much like its sequential implementation, the parallelized NAT enforces unique ports inside each core. It does not, however, enforce this uniqueness \emph{across} cores, a feature that does not break semantic equivalence. Whereas on the sequential implementation the allocated ports were used to distinguish between sessions, now the sharding solution allows for packets sent to different cores (hence pertaining to different external servers) to have the same allocated ports.

\subsubsection{CL}
A Connection Limiter (CL) aims to limit how many connections any single client (source IP) can make to any single server (destination IP) over a wider time frame (\eg several days).
Given the longer time frames involved, this NF uses a memory-efficient count-min sketch~\cite{count-min-sketch} to estimate the connection count from each client to each server.
For new connections, the source and destination IPs are used to index the sketch, indexing a configurable number of entries based on different hashes (5 by default in our case).
If all entries surpass the connection limit, the packet is dropped, preventing the new connection.
Otherwise, each entry is incremented.

As with the PSD, \maestro finds two different access patterns: the 5-tuple indexes a connection tracking map, while the source and destination IPs index the sketch. Again, the latter constraint subsumes the former and \maestro shards based on source and destination IPs.

\subsubsection{LB}
LB is a Maglev-like load balancer~\cite{eisenbud2016maglev}. Its main goal is to distribute traffic coming from the WAN to a series of identical servers on the LAN. LB registers new servers when it receives their packets coming from the LAN, and matches packets coming from the WAN with previously registered servers, keeping track of flows to ensure the same server handles packets from the same flow.

In order to maintain semantic equivalency between a shared-nothing parallel implementation and a sequential implementation, packets that find an available server in the sequential implementation must also find it available in the other.
This ultimately means that all cores would need to have all backends registered in their local state.
That said, packets coming in from the LAN in such a parallel implementation would only be able to be registered in a single core, preventing packets that arrive at other cores from seeing it.
With this limitation in mind, it becomes impossible for multiple cores to hold an identical set of backend servers without coordination, thus preventing the use of a shared-nothing model.
The \maestro analysis detects this issue when analyzing the LB SR.
Lacking a better alternative, \maestro issues a warning and opts for a read/write lock based approach.

\subsection{Performance Benchmarking Methodology}

To benchmark the NFs, we use a standard testbed topology~\cite{rfc2544}, connecting a traffic generator (TG) and a device under test (DUT), as shown in \cref{fig:testbed}. Both devices connect through a top-of-rack (TOR) switch from which we collect packet counters at the end of each experiment.
Both TG and DUT are equipped with dual socket Intel Xeon Gold 6226R @ 2.90GHz, 96~GB of DRAM, and Intel E810 100~Gbps NICs.
Turbo Boost, Hyper-Threading, and power saving features were disabled, as recommended by DPDK.

To measure throughput, the TG replays a given traffic sample (a PCAP file) in a loop at a given rate via the outbound cable for 10s per experiment.
The DUT receives this traffic, processes it, and sends it back via the return cable, allowing the TG to measure latency. We further use the TOR to infer loss at the DUT, and---through comparison with the TG report---to also detect when packets were lost within the TG as well.
We use DPDK-Pktgen~\cite{Pktgen} on the TG to find the maximum rate with less than $0.1\%$ loss.
We exclude and repeat sporadic experiment runs where loss within the TG---as opposed to the DUT---limited the results.
When studying scalability, we repeatedly reevaluate the NF, while varying the number of cores it may use.
We perform 10 measurements per experiment for statistical relevance and show error bars with min/max values. 
Our experiments properly handle NUMA considerations and indirection table rebalancing (\cref{section:implementation}).

\subsection{Picking the Workload}
\label{subsection:workload}

In this section we analyze how different workloads impact performance, and ultimately establish the right workload configuration to evaluate all \maestro's parallelization solutions.

\begin{figure}[t]
    \centering
    \includegraphics[width=\linewidth]{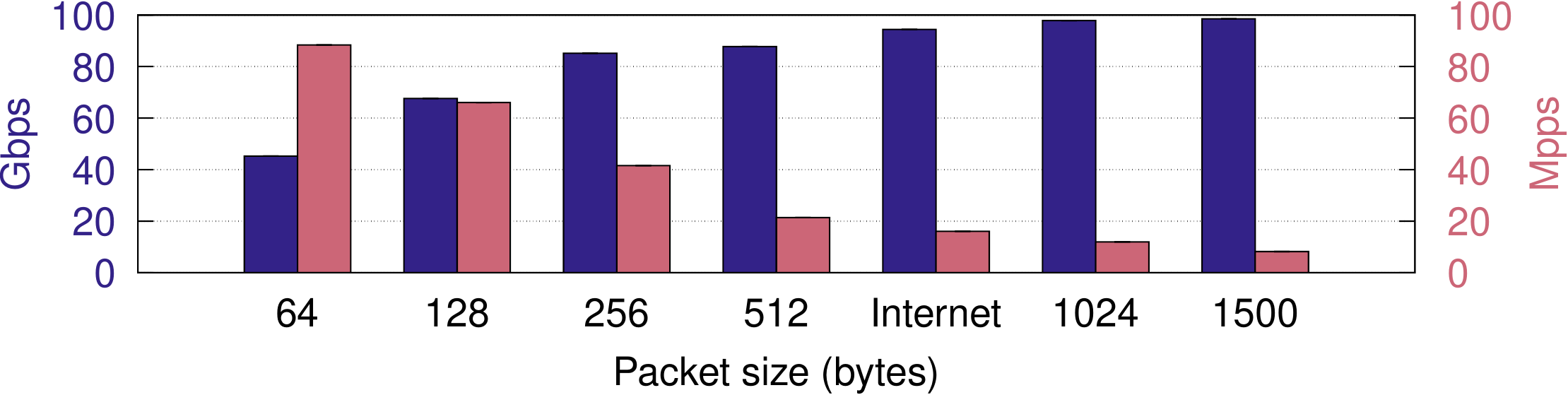}
    \caption{Throughput in Gbps (blue) and Mpps (red) of the parallel NOP running on 16 cores for different packet sizes.}
    \vspace{-2em}
    \label{fig:pkt_sz}
\end{figure}

\textbf{Packet size.} To measure the impact of packet size on the performance of NFs, we ran the NOP on all cores and generated traffic with fixed-sized packets (40k uniformly distributed flows), varying the size on each iteration. 
The results (\cref{fig:pkt_sz}) show that typical Internet traffic~\cite{benson2010network} and large packets easily achieve line-rate (100G), but that smaller packets struggle to keep up, reaching only \textasciitilde{}45Gbps with 64B packets---even with such a trivial NF.
Prior work~\cite{neugebauer2018understanding,agarwal2022understanding,enso} has pointed out that this bottleneck comes from PCIe 3.0 x16 and cannot be overcome without improved hardware.
Unless stated otherwise, further experiments in this paper use 64B packets.
As we measure more complex NFs that limit throughput below the 90Mpps shown in~\cref{fig:pkt_sz}, the bottleneck shifts from PCIe to the CPU, illustrating the NF's intrinsic performance.

\textbf{Churn.} The performance of parallel NFs can vary significantly for read or write workloads.
In networking terms, this typically relates to \emph{churn}, or the rate at which new flows are added and expired.
This is particularly important for lock and TM based implementations, where creating new flows can lead to costly aborted transactions or exclusive write locks.

We start by studying these churn effects on performance by focusing on the read/write lock-based parallel firewall, and comparing it to its shared-nothing counterpart.
To conduct churn experiments, ideally one would generate traffic live that changes flows periodically in an online manner.
We found it challenging to generate such traffic programmatically at line-rate so we followed an alternative solution: generating PCAPs with different levels of \emph{relative churn}---measured in $\text{flows} / \text{Gbit}$.
As Pktgen varies the replay rate of the PCAP to probe the NF, the resulting \emph{absolute churn}---measured in $\text{flows} / \text{minute}$ or fpm---changes in tandem.
This guarantees that our experiments converge to an equilibrium where the highest rate is found for the given churn.
Once we find this rate, we can multiply the PCAP's relative churn with the experimental rate to compute the absolute churn.

With this in mind, we built PCAPs which
(i) were small enough to fit in memory;
(ii) changed enough flows to produce the desired relative churn;
(iii) evenly spread these changes throughout the traffic;
and (iv) were cyclic (\ie the flows that expire at the start of the PCAP are created at the end).
We then replay these files in a loop for 10s as in all other experiments.

\begin{figure}[t]
    \centering
    \includegraphics[width=\linewidth]{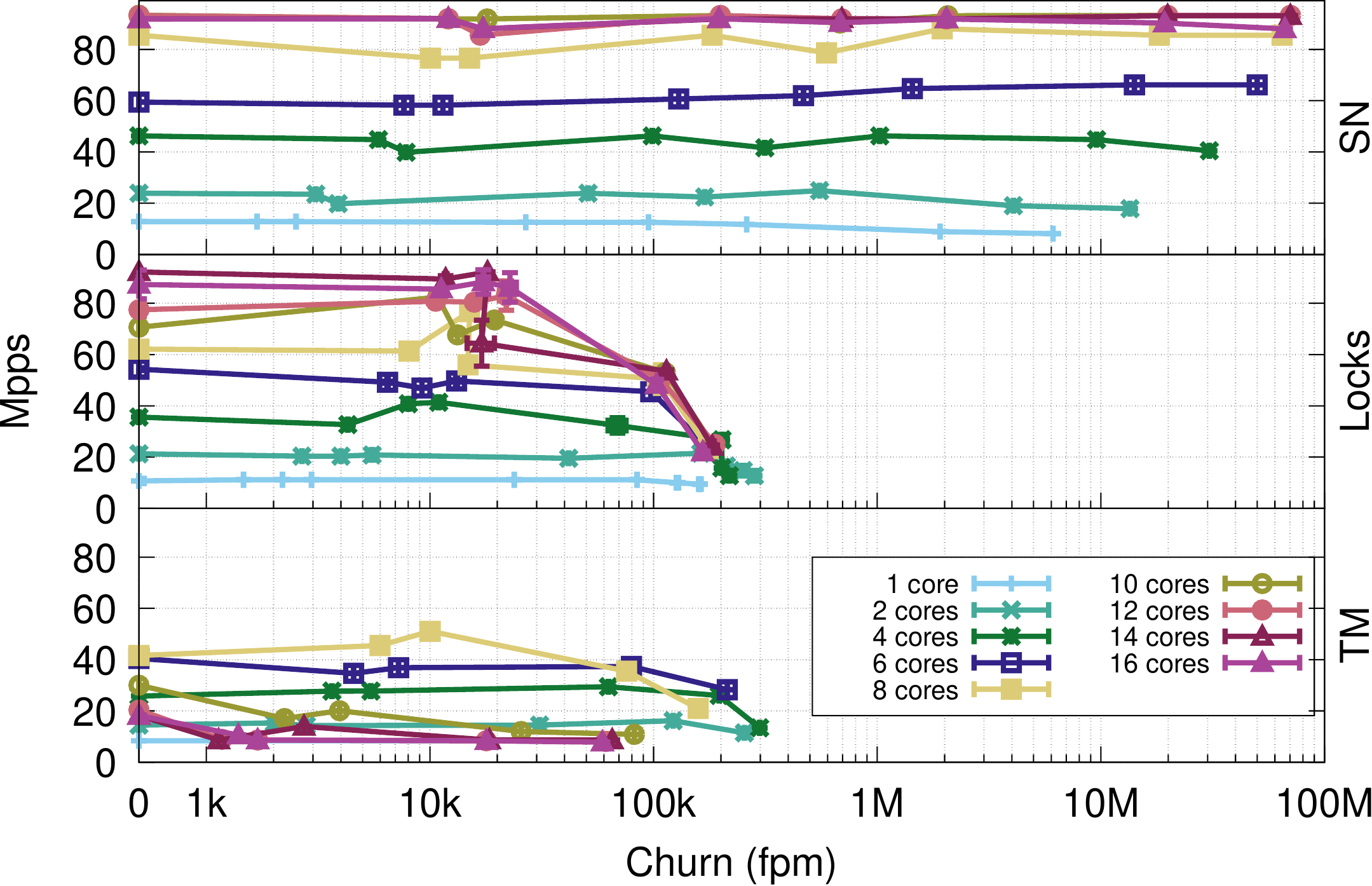}
    \caption{Churn study of the shared-nothing (top), lock-based (middle), and TM (bottom) parallel firewall.}
    \label{fig:churn}
    \vspace{-1em}
\end{figure}

\begin{figure}[t]
    \centering
    \includegraphics[width=0.98\linewidth]{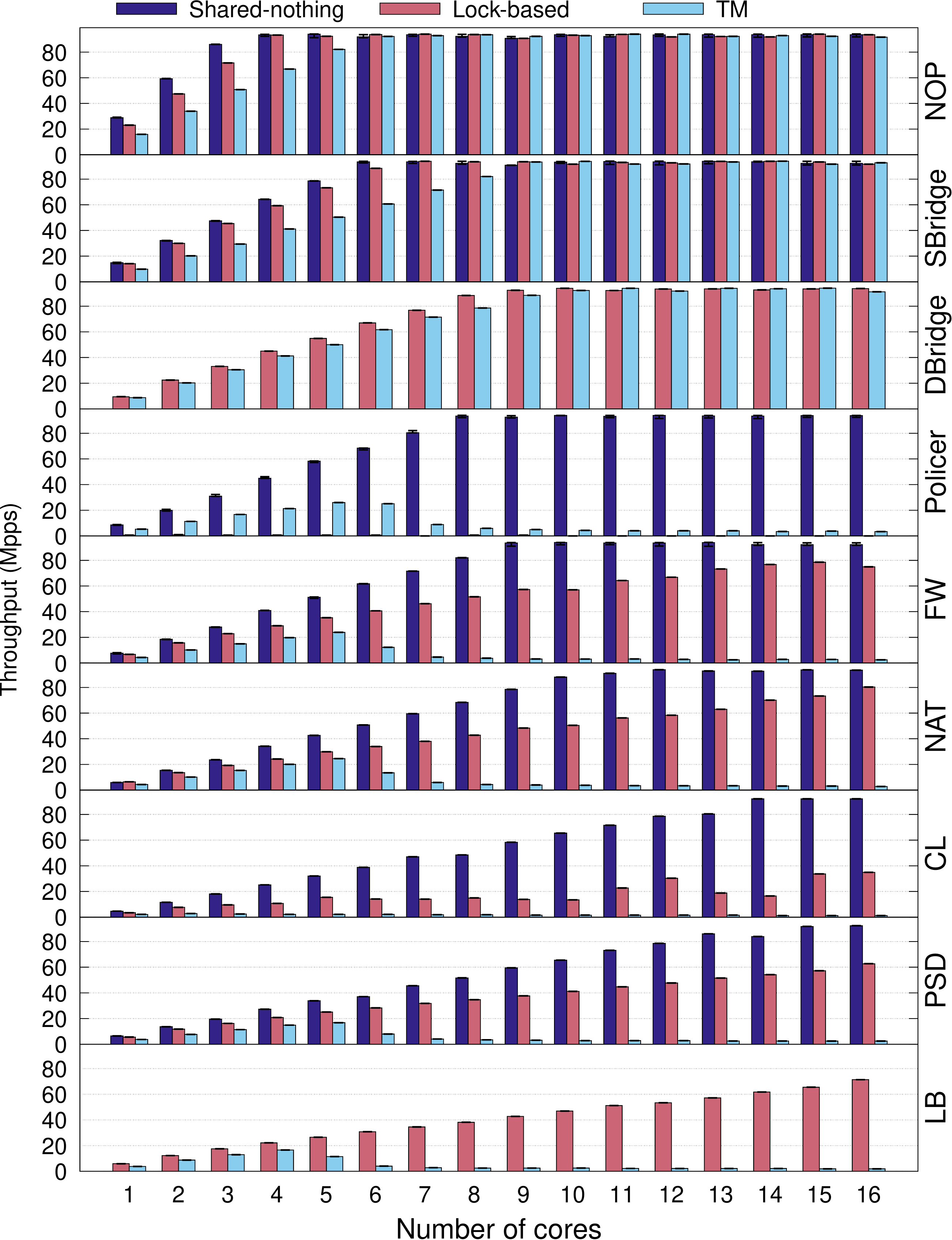}
    \caption{Parallel NF scalability with uniformly-distributed, read-heavy, small packets, using a shared-nothing approach when possible, read/write locks, and TM. \maestro cannot do a shared-nothing DBridge or LB.}
    \label{fig:technologies}
    \vspace{-1.5em}
\end{figure}

\cref{fig:churn} shows how the FW---parallelized with different approaches---scales under varying amounts of churn.
As absolute churn is computed based on the achieved rate, note that it too has error bars.
Under low or no churn, the lock-based FW scales well until bottlenecked by PCIe.
At a churn of \textasciitilde{}100k fpm we start observing the collapse of performance as the use of more cores just wastes more cycles busy-waiting under exclusive write locks.
Under heavy churn, performance is abysmal as all cores end up contending for write locks.
Note that the churn limit of an NF depends on the size of packets---\cref{fig:churn} uses 64B packets but for Internet traffic~\cite{benson2010network} the lock-based FW handles churn up to 400k fpm.

The results also show just how badly the FW parallelized with transactional memory handles churn. Although a useful tool in other domains, it proves ineffective when dealing with networked applications under churn.

The shared-nothing approach, unlike the lock-based one, suffers almost no performance variation with churn up to at least \textasciitilde{}100M fpm, a great advantage over the lock-based implementation.
Benson \etal~\cite{benson2010network} tell us to expect up to 6M fpm in typical data-center traffic---within the ability of our shared-nothing FW, but not the lock-based one.
University networks---typically with less than 15k fpm---could easily be handled even by our lock-based FW.

We focus the rest of this evaluation on studies without churn, giving the lock and TM based approaches the benefit of the doubt and illustrating their \emph{best-case} performance.

\subsection{Performance benchmarks}
\label{subsection:benchmarks}

With parallel versions of each of the above 8 NFs generated, we now evaluate their performance and scalability.
By default, \maestro generates a shared-nothing implementation when possible, falling back to read/write locks otherwise.
This choice can, however, be overriden, and \maestro can specifically generate parallel implementations using read/write locks and TM for any of the NFs, upon request.

\textbf{Parallelization technologies.} We now study the performance and scalability of each NF while being parallelized for each of the three approaches.
As per~\cref{subsection:workload}, the workload used is composed of uniformly-distributed, read-heavy, small packets\footnote{Experimental results using Zipfian traffic are shown in~\cref{appendix:zipf}}.
\cref{fig:technologies} shows throughput as a function of the number of cores.
Our raw performance is comparable to measurements from other recent works~\cite{farshin2021packetmill}, but we focus our attention on \emph{scalability}.
Though most NFs top out their performance before using all 16 cores due to bottlenecks in the PCIe bus or the memory controller, the takeaway here is the relative performance of the different approaches.

For all NFs where a shared-nothing approach was feasible, this option scales linearly until bottlenecked by the PCIe bus and then plateaus---an ideal outcome.
The lock-based implementations---though slower than their shared-nothing counterparts when available---still scale fairly well but do not always reach the PCIe bottleneck with 16 cores\protect\footnote{Eventually, all lock-based NFs except for the Policer and CL can reach the PCIe bottleneck using extra cores from the remote NUMA node.}.
The Policer shows what happens to these locks when writes are inevitable: as every packet must update the token bucket state, every packet requires an exclusive write lock, and performance suffers catastrophically.
Fortunately, this NF can be sharded by IP address, so is amenable to the shared-nothing approach.

The benefits of state sharding (\cref{subsec:sharding}) become clear when we compare the shared-nothing approaches with the lock-based ones for the more state intensive NFs, \ie the FW, NAT, CL, and PSD. When each core holds less state due to sharding, more of it fits in the core-local L1+L2 cache. In a shared-nothing approach where cores work independently on different working-sets this leads to an added performance improvement due to better caching, in addition to the benefits of parallelization. As a result, performance for few ($<4$) cores can be worse than linear scalability would predict and using many cores can have an added boost in comparison. Running these experiments with a workload of only 256 flows---which fits entirely in L1 cache---nullifies this effect.

A surprising takeaway is that TM  does not work well with the kinds of workloads found in more complex NFs, even in the absence of churn.
For simpler NFs it performs quite well, scaling linearly with the number of cores, though still operating more slowly than both shared-nothing and lock-based alternatives.
In these cases TM eventually catches up with the other approaches, albeit needing more cores to do so.
However, for more complex NFs TM performs abysmally, as the likelihood of a transaction aborting increases.

Ultimately, the clear winner is the shared-nothing approach, with the best backup option consistently being our read/write locks.
The PSD---our most CPU intensive NF which stands to gain the most from parallelization---performs $19\times$ better with 16 cores than a single-core version, due to the \emph{compound effects} of parallelization and improved cache efficiency.

\maestro does not deeply affect latency. We subjected all NFs to a 1Gbps uniform background traffic of 64B packets and collected 1000 latency probes within 10 seconds.
We detected no noticeable differences on the average and tail latency values between the sequential NFs and their respective parallel implementations, regardless of the adopted parallelization strategy.
Pktgen measured an average of $12 \pm 2 \mu s$ for CL and $11 \pm 1 \mu s$ for the remaining NFs.

\textbf{VPP comparison.} Finally, we compare \maestro with the Vector Packet Processing framework (VPP)~\cite{barach2018high,fdio-vpp}, which extends the concept of batch processing to the entire packet processing pipeline with the purpose of increasing performance by minimizing instruction cache misses.
VPP follows a converse approach to \maestro: packets are processed in batches in a shared-memory parallel environment where packets can end-up on any core without regard to flows or locality.
Developers must then adapt the way they implement the NF to those assumptions.
This approach can require more expertise and development effort, but once NFs are built in this way the framework handles many of the low-level details.

To compare the performance of a \maestro parallelized NF with an expertly developed one for VPP, we pitch our NAT against the VPP \texttt{nat44-ei} with the DPDK plugin.
These two NFs are the most similar we found between the VPP distribution and our corpus. We further removed a number of features from the VPP NAT to bring their implementations even closer together\protect\footnote{We removed statistical counters, disabled IPv4 checksum checking, completely removed the IPv4 reassembly feature, and finally replaced the IPv4 lookup with static forwarding.}.

\begin{figure}[t]
    \centering
    \includegraphics[width=\linewidth]{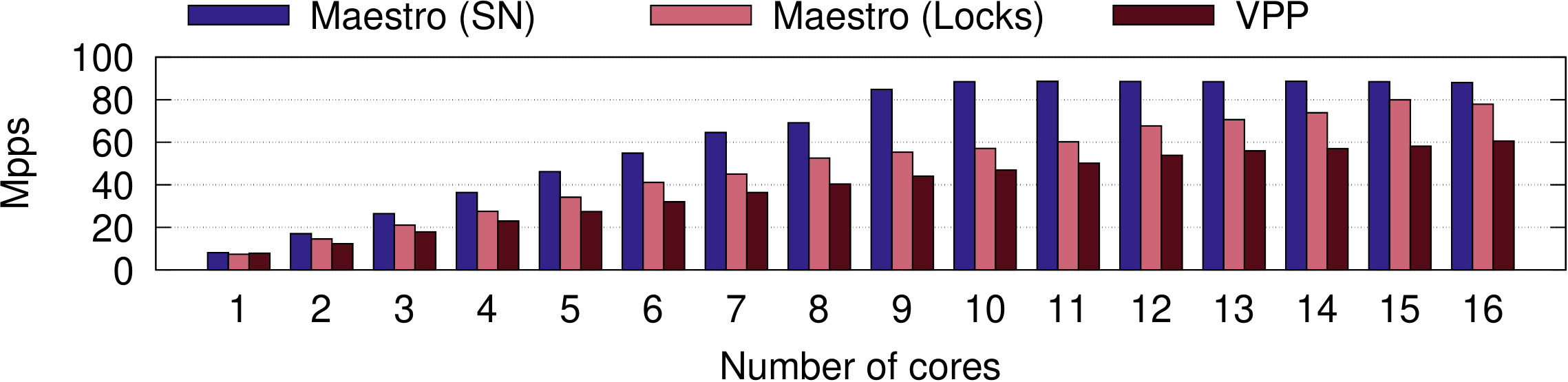}
    \caption{VPP and \maestro NAT comparison.}
    \label{fig:vpp-64b}
    \vspace{-2em}
\end{figure}

\cref{fig:vpp-64b} shows the performance comparison between the parallel \maestro NAT (shared-nothing and lock-based) and \texttt{nat44-ei}, all under uniformly distributed 64B packets.
Though all approaches scale well, \maestro's shared-nothing decisively outperforms VPP, reaching the PCIe bottleneck with 10 cores.
This is due to the shared-memory design that VPP follows.
A fairer comparison would be between VPP and the lock-based \maestro NAT, as both use shared-memory.
Here both scale more slowly, never fully reaching the PCIe bottleneck up to 16 cores.
\maestro slightly outperforms VPP. Further investigation with the \texttt{perf}~\cite{perf} tool showed us that although the \maestro lock-based NAT and the VPP one perform very similar numbers of memory reads and writes per packet, the \maestro NAT more frequently finds the data on L1 cache (\maestro's $55\%$ vs.\ VPP's $46\%$) and has to access RAM less frequently (\maestro's $3\%$ vs.\ VPP's $4\%$).
The key takeaway though, is that \maestro's \emph{automatically} parallelized NFs perform competitively with expertly developed, manually parallelized NFs, without as much of a hassle.

%% file: sections/7-related-work.tex
\section{Related Work}
\label{section:rw}

\noindent \textbf{Fast packet processing.} To address the performance challenges associated with software NFs, new packet I/O frameworks were proposed \cite{Bonelli2016,Rizzo2012,Deri2004,ebpf}. To achieve high packet processing rates these solutions explore several types of optimizations including zero-copy, kernel bypass, I/O batching, and multi-queue support~\cite{Barbette2015}. VPP~\cite{barach2018high} even expands batching to the whole packet processing pipeline in order to reduce instruction cache misses.
Most implementations of network functions today~\cite{Trevisan2016,Emmerich2015,Wu2018,Zaostrovnykh2017}, including those from \maestro, rely on Intel DPDK~\cite{Intel2010}, a kernel-bypass packet processing framework that provides a set of software libraries and drivers for fast packet processing, providing multi-core and NUMA-aware functionalities.

\noindent \textbf{NF acceleration.} PacketMill~\cite{farshin2021packetmill} accelerates NFs by carefully managing packet metadata and performing code-optimizations across the whole network stack. Another approach to improve the performance of a software NF is to leverage the platform hardware. Previous work~\cite{Papadogiannaki2017, Trevisan2016,Jamshed2012,Dobrescu2009,maple} has explored multi-core CPU architectures, showing the significant improvements they can achieve on an NF's performance, but also the challenges involved. Papadogiannaki \etal {\cite{Papadogiannaki2017}}, for instance, explored the advantages of a shared-nothing model over a lock based implementation.
The goal of \maestro is to offer the advantages of parallelization to NFs, for free.
Although their work focused on the most efficient utilization of available resources, we use their shared-nothing model as guidance for automated generation of parallel network functions.
These solutions are \emph{manual}, requiring extensive expertise and painstaking effort from the developer.

De Carli \etal~\cite{de2014beyond} proposed a concurrency model for software IDSes that uses program analysis to infer the NF's flow semantics, feeding that information to a software scheduler that steers packets to shared-nothing threads. Though the concepts share similarities, \maestro's approach differs from theirs by (1) considering a wider class of NFs more generally, rather than IDSes in particular; (2) using ESE to extract fine-grained state access patterns, as opposed to their less granular program-slicing approach; and (3) handling packet steering entirely in hardware by generating RSS configurations for NICs, avoiding the bottleneck of the software scheduler and allowing \maestro parallelized NFs to scale better.

\noindent \textbf{Flow steering.} Although some NICs support rich flow-steering configurable features~\cite{e810,connectx5}, these are orthogonal to RSS and do not replace it. Using them to assure semantic equivalence on a shared-nothing implementation may require frequently adding/deleting a large amount of rules (specially under high churn), which can heavily affect performance~\cite{katsikas2021}.

\noindent \textbf{NF verification and synthesis.} In recent years, verification techniques have started to be applied to network functions. Some of the most relevant work includes verification of network properties \cite{kazemian2012header,khurshid2013veriflow}, configurations \cite{fogel2015batfish,beckett2017general}, and NFs \cite{zaostrovnykh2019verifying}. More recently, the research community has started exploring synthesis approaches for SDN-based control~\cite{chen2018towards}, data plane programs~\cite{gao2020switch, zhang2020gallium, synapse}, and BGP configurations~\cite{vanbever2020,vanbever2021}.
Our work fits into this line, by analyzing sequential NFs to automatically generate accelerated versions.

%% file: sections/8-conclusions.tex
\section{Conclusions}
\label{section:conclusions}

In this paper we presented \maestro, a tool to automatically parallelize sequential network functions. \maestro judiciously configures the NIC's RSS mechanism to distribute traffic across cores, while preserving semantics, resorting to locking mechanisms only when necessary. \maestro significantly improved performance for all the NFs we analyzed---scaling-up performance linearly until hitting fundamental bottlenecks in PCIe, the memory controller, or line-rate---while reducing developer effort to the push of a button.

%% file: sections/9-acknowledgments.tex
\section*{Acknowledgments}

We are grateful to our shepherd, Tom Barbette, and the anonymous NSDI'24 reviewers. We thank Hugo Sadok for his comments on earlier drafts of the paper. We also thank Paolo Romano and Daniel Castro for their help on handling the TM approaches. This work was supported by the European Union (ACES project, 101093126), INESC-ID (via UIDB/50021/2020), and the SALAD-Nets CMU-Portugal/FCT project (2022.15622.CMU). Francisco Pereira was supported by the FCT scholarship PRT/\allowbreak BD/\allowbreak 152195/\allowbreak 2021.


%% file: sections/10-appendix.tex
\clearpage
\appendix

\section{Appendix}

\subsection{Code excerpts from \maestro}
\label{appendix:code}

\begin{figure}[b!]
    \centering
    \includegraphics[width=\linewidth]{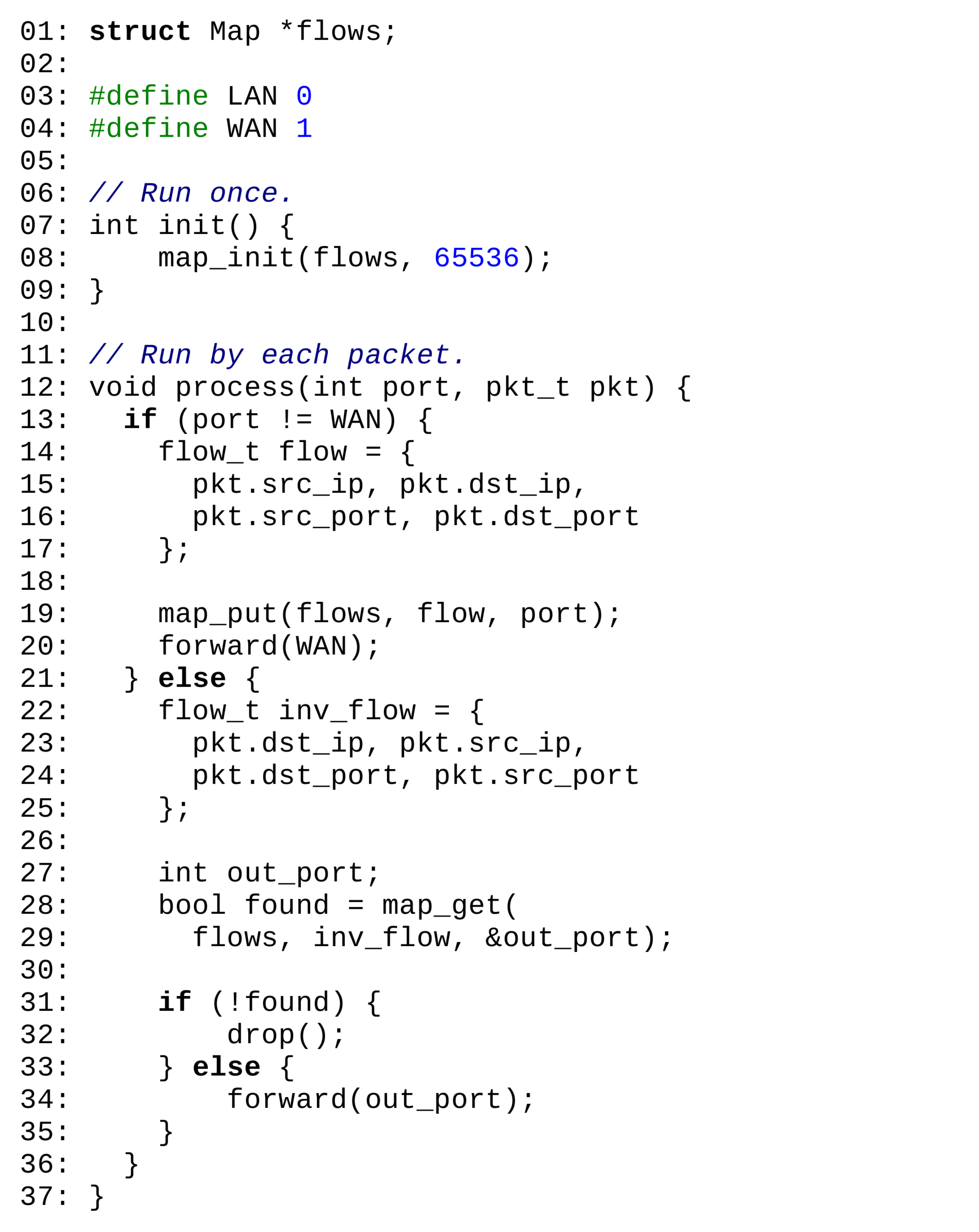}
    \caption{Pseudo-code of the sequential firewall used as an example throughout the paper.}
    \label{fig:fw-seq}
\end{figure}

\begin{figure}[h!]
    \centering
    \includegraphics[width=\linewidth]{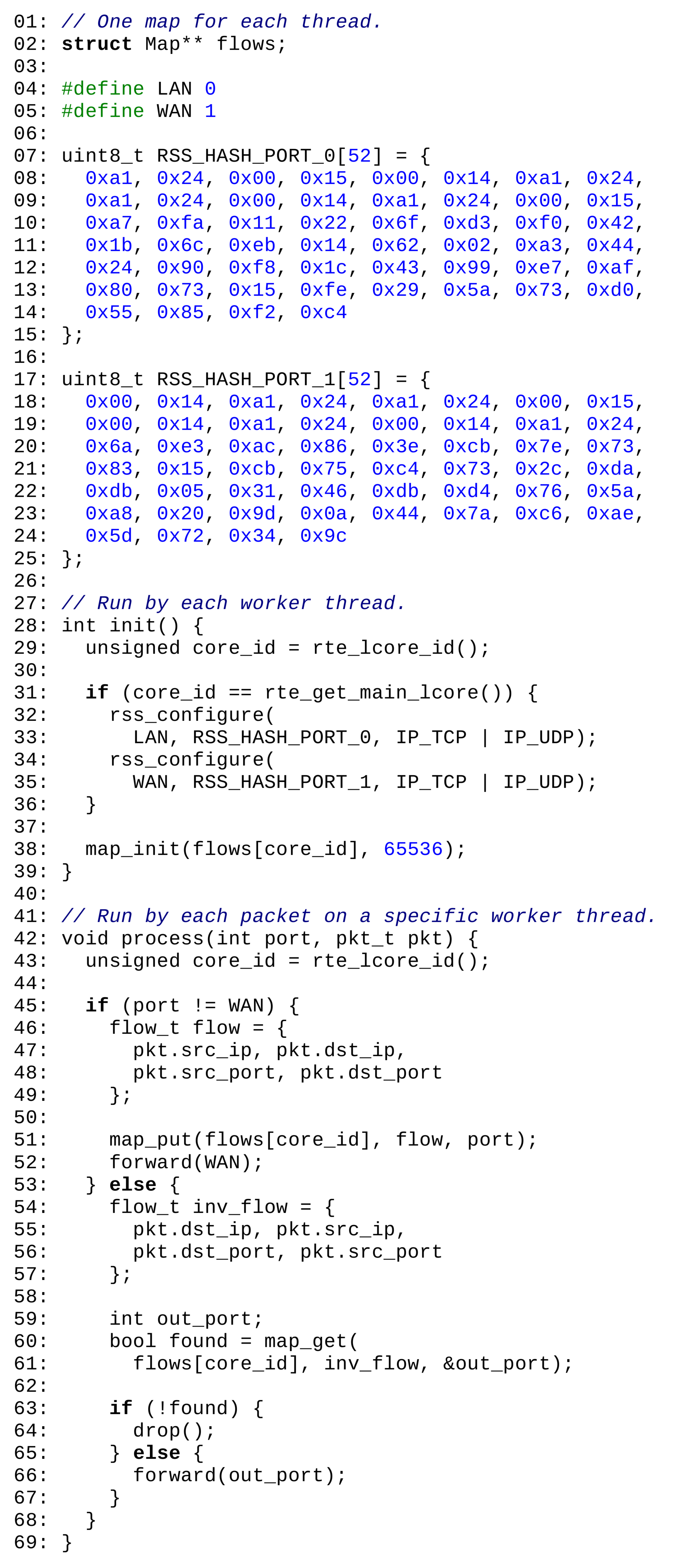}
    \caption{Pseudo-code of the firewall, but now parallelized by \maestro with a shared-nothing architecture (and described in~\cref{subsubsection:microbenchmarks:fw}).}
    \label{fig:fw-sn}
\end{figure}

We present here the pseudo-code of the firewall NF used throughout the paper, both its sequential and parallel shared-nothing implementations. These serve to provide a sense of what the \maestro pipeline both accepts as input (\cref{fig:fw-seq}) and automatically generates as output (\cref{fig:fw-sn}). As such, we reiterate that these are \emph{not} complete examples, but only pseudo-code, as they were shortened and simplified for clarity purposes. The complete solutions can be found on our public GitHub repository~\cite{maestro-repo}.

Notice the symmetry of the RSS hashes (lines 7 to 25 in~\cref{fig:fw-sn}), as it is what ultimately enables its shared-nothing approach. As explained in \cref{subsection:microbenchmarks}, this symmetry allows packets coming from the WAN to be sent to the same core as their corresponding symmetric packets from the LAN.


\subsection{Macrobenchmarks with Zipfian traffic}
\label{appendix:zipf}

\begin{figure}[t]
    \centering
    \includegraphics[width=\linewidth]{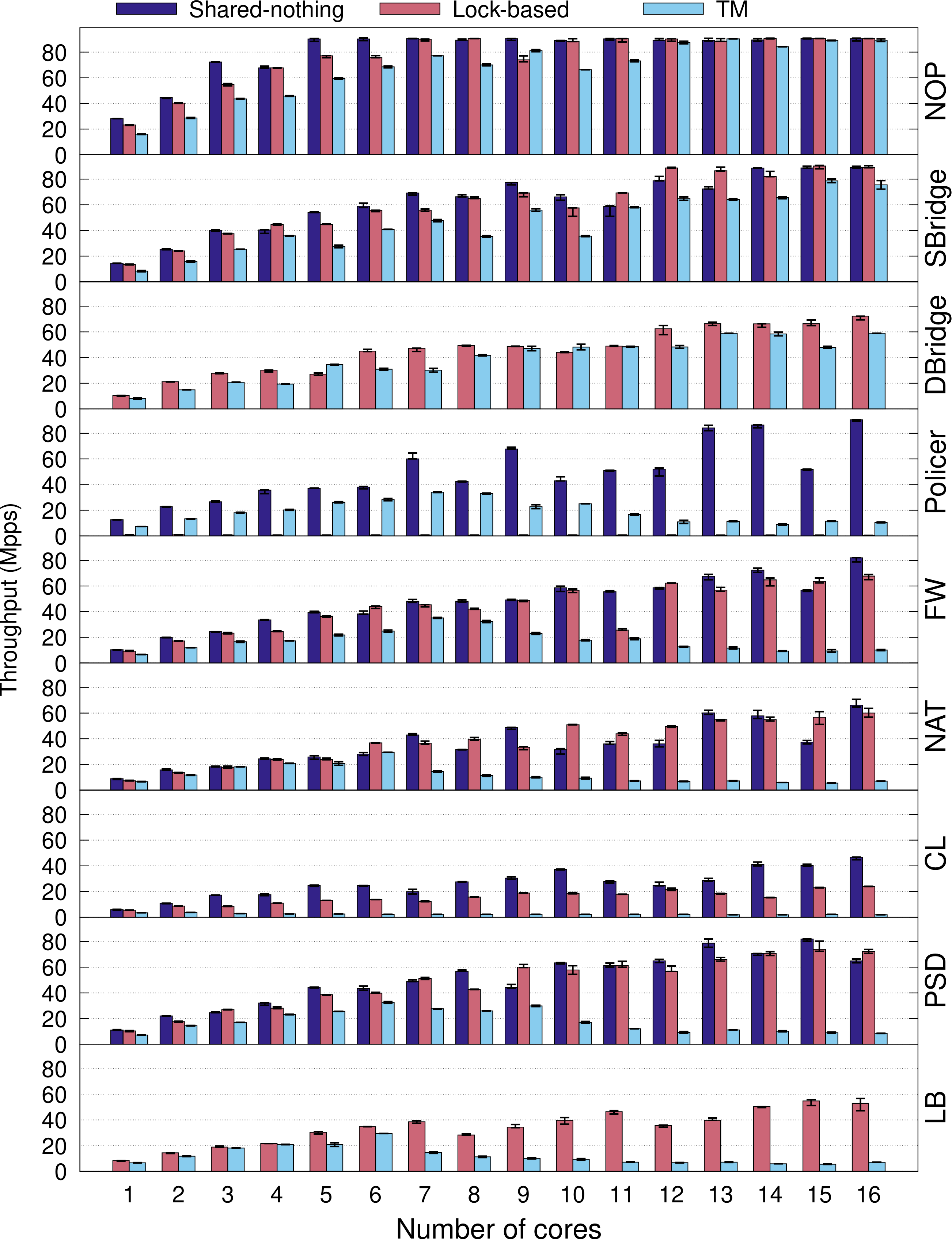}
    \caption{Parallel NF implementation scalability with Zipfian, read-heavy, small packet traffic, using a shared-nothing approach when possible, read/write locks, and TM.}
    \label{fig:technologies-zipf}
\end{figure}

While in \cref{fig:technologies} we show how throughput varies for different parallelization techniques under uniform traffic, here we repeat the experiment with Zipfian traffic instead~\cite{benson2010network} (we describe this Zipfian traffic in~\cref{section:implementation}). We balanced the indirection table for each implementation to better handle the skew, as described in~\cref{subsec:skew}. The results are shown in~\cref{fig:technologies-zipf}.

The key takeaways are the same as in~\cref{fig:technologies}: when available the shared-nothing approach is always preferred; the lock-based solutions frequently do not scale as well as their shared-nothing alternatives and suffer in more state-intensive NFs; and TM-based approaches perform unreliably.

We do, however, find differences between these results and their counterparts under uniform traffic. Although under uniform traffic it is rather clear that throughput scales up with the number of cores when using the shared-nothing approach, with Zipfian traffic this scaling is not always consistently monotonic. This is to be expected, as the efficacy of balancing load across cores may not consistently improve when more cores are added. Indeed, when many cores are used, a single elephant flow can bottleneck a single core, limiting the maximum throughput we will be able to achieve in our experimental setup. This is particularly limiting for computationally and state intensive NFs---such as the the Connection Limiter--- which are unable to perform as well with Zipfian traffic as they do with uniform. These results nevertheless confirm that \maestro generated NFs almost always perform as well with Zipfian traffic as they do with uniform.

\subsection{Reproducibility}

We make \maestro's code publicly available in~\cite{maestro-repo}. In that repository, one can find not only the source code for the entire pipeline, but also the complete set of NFs we mention on this paper, along with their corresponding parallel solutions found by \maestro and described in~\cref{subsection:microbenchmarks}.

We also make available our test suit in~\cite{maestro-test-suit}. It contains all the required scripts to generate~\cref{fig:traffic,fig:pkt_sz,fig:churn,fig:technologies,fig:vpp-64b,fig:technologies-zipf}. They were tested on 2 machines with dual socket Intel Xeon Gold 6226R @ 2.90GHz, 96~GB of DRAM, and e810 Intel NICs~\cite{e810}, and running Ubuntu 22.04.
